\documentclass[amsmath,amssymb]{article}

\usepackage{mathtools,cancel}

\usepackage{soul,xcolor}
\usepackage{xcolor}
\usepackage{multirow}
\usepackage{pstricks}
\usepackage{dcolumn}
\usepackage{bm}
\usepackage{latexsym}
\usepackage{dcolumn}
\usepackage[utf8x]{inputenc} 
\usepackage{amsmath}
\usepackage{amsfonts,amssymb}
\usepackage{graphicx,epsfig}
\usepackage{color}
\usepackage{psfrag}
\usepackage{amsthm}
\usepackage{ulem} 
\usepackage{soul} 
\usepackage{flushend}
\usepackage{titlesec}
\usepackage{slashed}
\usepackage{authblk}
\usepackage[font=footnotesize]{caption}
\usepackage[hmarginratio=1:1,top=32mm,columnsep=60pt]{geometry}

\newcommand{\bea}{\begin{eqnarray}}
\newcommand{\eea}{\end{eqnarray}}
\newcommand{\be}{\begin{equation}}
\newcommand{\ee}{\end{equation}}
\newcommand{\ba}{\begin{array}}
\newcommand{\ea}{\end{array}}

\begin{document}

\unitlength = 1mm
\begin{flushright}
SHIP-HEP-2020-03
\end{flushright}

\title{\textbf{Searching for charged lepton flavor violation \\ at $ep$ colliders}}

\author{\normalsize Stefan Antusch$^{\dagger}$, A. Hammad$^{\dagger\delta}$ and Ahmed Rashed$^{\ddagger}$}
\affil{\small
$^\dagger$Department of Physics, University of Basel, Klingelbergstr.\ 82, CH-4056 Basel, Switzerland\\
$^{\ddagger}$ Department  of Physics,  Shippensburg University of Pennsylvania,\\
 Franklin Science Center, 1871 Old Main Drive, Pennsylvania, 17257, USA\\
 $^{\delta}$ Centre for theoretical physics, the British University in Egypt, P.O. Box 43, Cairo 11837, Egypt
 }
\date{}

{\let\newpage\relax\maketitle}

\begin{abstract}
\noindent \normalsize We investigate the sensitivity of electron-proton ($ep$) colliders for charged lepton flavor violation (cLFV) in an effective theory approach, considering a general effective Lagrangian for the conversion of an electron into a muon or a tau via the effective coupling to a neutral gauge boson or a neutral scalar field. For the photon, the $Z$ boson and the Higgs particle of the Standard Model, we present the sensitivities of the LHeC for the coefficients of the effective operators, calculated from an analysis at the reconstructed level. As an example model where such flavor changing neutral current (FCNC) operators are generated at loop level, we consider the extension of the Standard Model by sterile neutrinos. We show that the LHeC could already probe the LFV conversion of an electron into a muon beyond the current experimental bounds, and could reach more than an order of magnitude higher sensitivity than the present limits for LFV conversion of an electron into a tau. We discuss that the high sensitivities are possible because the converted charged lepton is dominantly emitted in the backward direction, enabling an efficient separation of the signal from the background.
 
\end{abstract}


\newpage

\section{Introduction}\label{intro}

Experimental searches for charged lepton flavor violation (cLFV) are among the most sensitive probes of new physics beyond the Standard Model (SM) of elementary particles. In the SM, such flavor changing neutral current interactions in the lepton sector are absent at tree level and with massless neutrinos, and even when neutrino masses are introduced in an effective theory approach via the dimension five neutrino mass operator, they only get induced at loop level at tiny rates far below envisioned observational possibilities. 

As an indirect probe of new physics, cLFV is known to be sensitive to extensions of the SM at scales far beyond the reach of direct searches at present and currently discussed future colliders. At present, particularly strong limits on LFV $\mu - e$ transitions come from $Br(\mu\to e\gamma)\le 4.2\times 10^{-13}$  \cite{TheMEG:2016wtm}, and on LFV $\tau - e$ transitions from $Br(\tau\to e\gamma)\le 3.3\times 10^{-8}$ \cite{Aubert:2009ag} and $Br(\tau\to 3e)\le 2.7\times 10^{-8}$  \cite{Hayasaka:2010np}. 
Planned experiments to extend the cLFV searches beyond these limits include  MEG II \cite{Baldini:2018nnn}, which could reach a sensitivity for $Br(\mu\to e\gamma)$ down to $6\times 10^{-14}$. Furthermore, the Mu3e experiment plans to reach a sensitivity for $Br(\mu\to 3e)$ down to $2\times 10^{-15}$ \cite{Arndt:2020obb}. Regarding muon to electron conversion, the Mu2e and COMET experiments have the goal to increase the sensitivity to the $\mu-e$ conversion rate by four orders of magnitude down to $3\times 10^{-17}$ \cite{Miscetti:2020gkk,Shoukavy:2019ydh}, and the PRISM project even aims at a sensitivity down to $10^{-18}$ \cite{Alekou:2013eta}. Both $B$-factories BABAR and  BELLE II aim to improve the sensitivity on LFV $\tau$ decays by more than an order of magnitude down to $Br(\tau\to e\gamma) < 3\times 10^{-9} $ and $Br(\tau\to 3e) < 1.2\times 10^{-9} $ \cite{Kou:2018nap,Bona:2007qt,Lusiani:2010eg}.

In this paper, we show that future electron-proton ($ep$) colliders such as the LHeC would be excellent facilities for probing the cLFV conversion of an electron into a muon or a tau via the effective coupling to a neutral gauge boson or a neutral scalar. To explore the potential for discovering cLFV induced by heavy new physics in  a model-independent way, we consider a general effective Lagrangian for our sensitivity calculations via collider simulations at the reconstructed level. In addition, as an example model where flavor changing neutral current (FCNC) operators inducing cLFV are generated at loop level, we consider the extension of the Standard Model by sterile neutrinos. There we show that the LHeC could probe the LFV conversion of an electron into a muon beyond the current experimental bounds, and could reach more than an order of magnitude higher sensitivity for the LFV conversion of an electron into a tau.

\section{High sensitivity to cLFV at $ep$ colliders} 
\label{sec.2}

Compared to electron-positron colliders, the high center-of-mass energy at $ep$ colliders can provide an environment to test the SM at high energies with comparably low rates of background. Two examples of possible future $ep$ colliders are the Large Hadron electron Collider (LHeC) \cite{Agostini:2020fmq,Bruening:2013bga,AbelleiraFernandez:2012cc,Klein:2009qt} and the $ep$ mode of the Future Circular Collider (FCC). At the LHeC, the center-of-mass energy of 1.3 TeV with a total of 3 $\text{ab}^{-1}$ integrated luminosity would be achieved by the use of the 7 TeV proton beam of the LHC in addition to a 60 GeV electron beam with up to $80\% $ polarization. Moreover, the proposed electron-proton experiment at the FCC (FCC-eh) is designed with the same energy level of the electron beam from the LHeC electron linac, but with the upgraded proton beam with energy of 50 TeV from the FCC-hh. This will achieve a center of mass energy of 3.5 TeV. This environment can be employed for significantly improving the PDF measurements and lower the associated systematic uncertainties. At the same time, an impact on the precision of some Higgs measurements is anticipated. In general, electron-proton colliders would be a great environment for testing certain types of new physics beyond the Standard Model, as has been explored in various studies (cf.\ e.g.\ \cite{Antusch:2020fyz,Jana:2019tdm,Flores-Sanchez:2019jcx,Azuelos:2019bwg,Antusch:2019eiz,DelleRose:2018ndz,Dev:2019hev}).

\subsection{cLFV via effective vertices at $ep$ colliders} 
\begin{figure}[h!]
\centering
\includegraphics[scale=0.3]{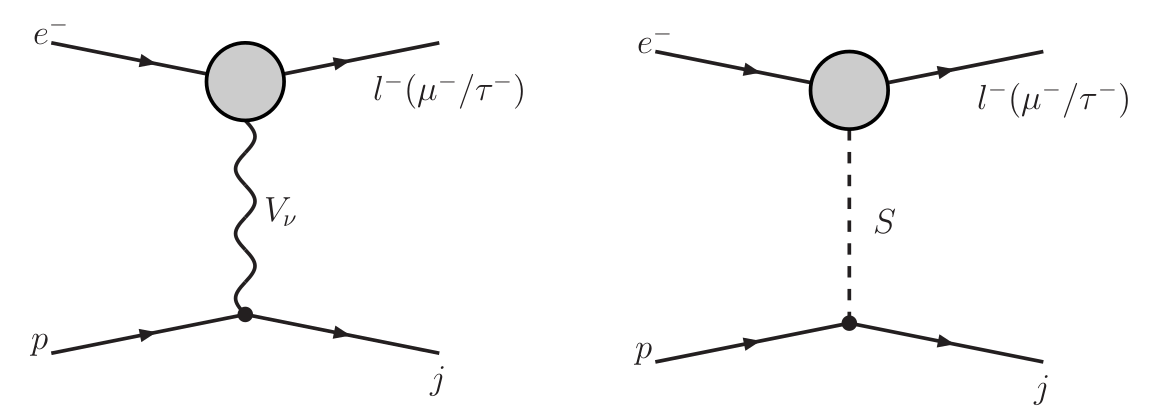}
\caption{Feynman diagrams for cLFV processes at the LHeC induced by effective operators (represented by blobs in the diagrams) that couple the incoming electron to a muon or a tau and a vector bosons $V_\nu$ (left) or a scalar $S$ (right).
}
\label{fig:1}
\end{figure}

Charged lepton flavor violating (cLFV) processes can occur at the LHeC through an effective vertex that couples the incoming electron to a muon or a tau and a neutral scalar or vector boson. With the neutral scalar or vector boson in the $t$-channel, the effective flavor changing neutral current (FCNC) interactions can lead to $e-\mu$ or $e-\tau$ flavor transitions, as shown in Fig.\ \ref{fig:1}. The processes have a specific kinematics that can be used to efficiently discriminate the signal from the SM background. A particularly useful feature, as we will discuss below in section \ref{sec:kinematics}, is that at low momentum transfer the final state lepton, i.e.\ the $\mu$ or $\tau$, is dominantly emitted in the backward region of the detector (cf.\ \cite{Agostini:2020fmq,AbelleiraFernandez:2012cc}). At the LHeC, we will show that this allows to almost completely suppress the relevant SM backgrounds in some cases.  

The effective FCNC Lagrangian for charged leptons contains effective operators coupling the charged leptons to neutral scalars and neutral vector bosons. The effective Lagrangian for the couplings to neutral scalars is given by 
\begin{align}
{\mathcal{L}^{\text{scalar}}_{\mathrm{eff}}} = \bar{\ell}_\alpha P_{L,R} \ell_\beta S\ N_{L,R}  ,
\label{eq:1}
\end{align}
with $\ell_\beta, \ell_\alpha, S$ representing the incoming and outgoing charged leptons and the neutral scalar boson of the effective vertex, respectively. $N_{L,R}$ represents the left and right form factors of the effective scalar operator and $P_{L,R}$ are the chiral projection operators. 
We note that expressions like $P_{L,R}  N_{L,R}$ are shorthand notations for the sum over both combinations, $P_{L}  N_{L}+P_{R}  N_{R}$. 
The part of the effective Lagrangian for the coupling to  vector bosons can be expressed in terms of monopole and dipole operators. The effective Lagrangian containing the monopole operators is given by 
\begin{align}
{\mathcal{L}^{\text{monopole}}_{\mathrm{eff}}} = \bar{\ell}_\alpha \gamma_\mu P_{L,R} \ell_\beta\left[A_{L,R}\ g^{\mu\nu}+ B_{L,R}(g^{\mu\nu} q^2-q^\mu q^\nu) \right] V_\nu ,
\label{eq:2}
\end{align}
where $q$ is the momentum of the gauge boson $V_\nu$ and where in the SM $V_\nu$ is either $Z$ or $\gamma$. $A_{L,R}$ and $B_{L,R}$ are the form factors of the monopole operators. The effective Lagrangian containing the dipole operator is given by 
\begin{align}
{\mathcal{L}^{\text{dipole}}_{\mathrm{eff}}} = \bar{\ell}_\alpha \sigma^{\mu\nu} P_{L,R} \ell_\beta\ q_\mu V_\nu\  D_{L,R} ,
\label{eq:3}
\end{align}
with $\sigma^{\mu\nu}=\frac{i}{2}[\gamma^\mu,\gamma^\nu]$ and $D_{L,R}$ denoting the left and right form factors of the dipole operator.

\subsection{Low background for cLFV due to specific kinematics}\label{sec:kinematics}
The differential cross sections of the cLFV processes (cf.\ Fig.\ \ref{fig:1}) depend on the center of mass energy $s$ and the two kinematic variables $q^2$ and the Bjorken variable $x$. At the electron-proton colliders, the Bjorken $x$ can be obtained from the measurement of the inelasticity $y_e$ as \cite{Klein:2008di}
\begin{align}
x = \frac{q^2}{s\ y_e},\hspace{6mm} \text{with}\hspace{6mm} y_e = 1- \frac{E_\mu}{2E_e}\left(1-\cos\theta \right),
\end{align}
with $E_\mu,E_e$ being the energies of the scattered muon and the incoming electron, respectively. The scattering angle $\theta$ is defined between the direction of the outgoing particles and the proton beam.  For the region of the parameter space with $x \approx E_e/E_p$,  
the  energy  of  the  scattered  muon  is  approximately equal to the electron beam, which causes the cross section to peak in the 
backward direction of the detector. For larger $q^2$, $x$ is larger due to the larger energy transfer from the proton beam that
 pushes the scattered muons somewhat more in the forward direction \cite{AbelleiraFernandez:2012cc}. 
\begin{figure}
\centering
~~~~~~\includegraphics[scale=0.5]{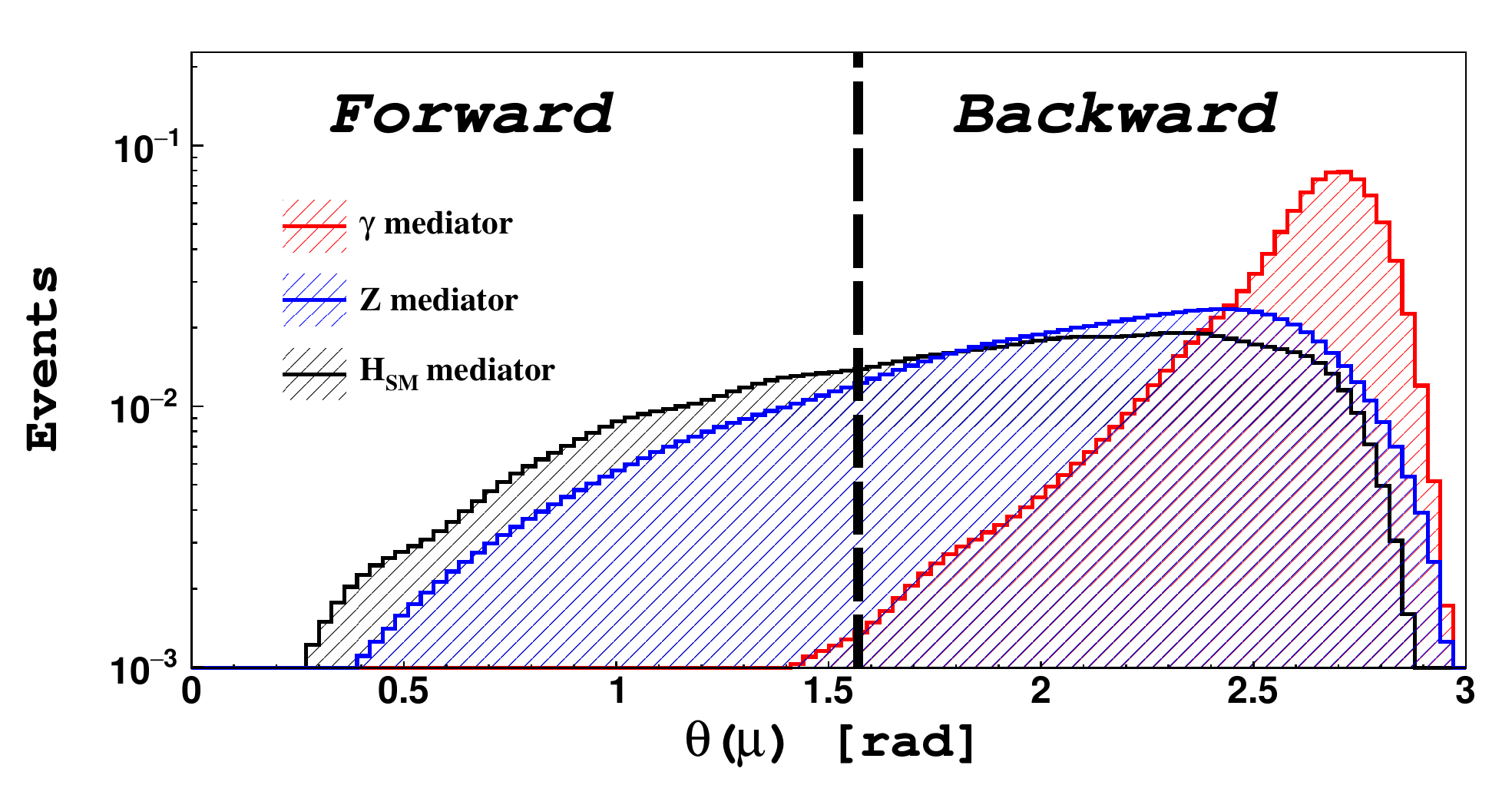}
\caption{Examples for the muon angular distributions at the reconstructed level for the photon (red), $Z$ boson (blue) and Higgs boson (black) mediated cLFV processes shown in Fig.\ \ref{fig:1}. The distributions in the plot correspond to the contributions from the form factors $A^{Z/\gamma}_{L,R}$ and $N^{H}_{L,R}$, with total number of events normalized to one. Note that the $y$-axis has a logarithmic scale. The forward direction is the proton beam direction and the backward direction is the electron beam direction. }
\label{fig:2}
\end{figure}
 
The SM background processes take place through the charged and neutral currents with $W^\pm$ and $Z/H$ bosons exchange. For the charged current, a ($t$-channel) $W$ boson can radiate a $Z/\gamma^\ast$ which then generates a $\ell\bar{\ell}$ pair. For the neutral currents, a ($t$-channel) $Z/H$ boson can generate charged leptons via radiating weak gauge bosons which then decay leptonically.
Other backgrounds come from the decay of the on-shell produced bosons, e.g. $pe-\to Z e^- j, Z\to\mu^\pm\mu^\mp$. The production of the on-shell $Z$ boson requires a large energy transfer, and thus the dimuons will be detected mainly in the forward region of the detector. Accordingly, the cLFV process at the LHeC through an effective vertex can provide a unique signal in the backward direction which is almost background free.    

In Fig.\ \ref{fig:2}, we show examples for the angular distribution of the scattered muons at the LHeC, for the case of  exchanged photons, $Z$ bosons, and SM Higgs particles (showing as examples the form factors $A^{Z/\gamma}_{L,R}$ and $N^{H}_{L,R}$). 
As one can see, the scattered muons are dominantly emitted in the backward direction. 
For the massive mediators $(Z, H)$, the cross section maximizes at $q^2=M^2$ and thus the peak shifts towards the forward direction compared to the photon case. A similar effect occurs for the form factors with momentum dependence, $B^{Z/\gamma}_{L,R}$ and $D^{Z/\gamma}_{L,R}$, which will be discussed in section 4 (with angular distributions shown in Fig.\ 7). 

For the simulation, we have implemented the effective vertices in MadGraph \cite{Alwall:2014hca}.
After generating the events by MadGraph, Pythia \cite{Sjostrand:2006za} is used for showering  and hadronization. For fast LHeC detector simulation we use Delphes \cite{deFavereau:2013fsa}. The event reconstruction has been done by MadAnalysis5 \cite{Conte:2012fm} with the requirement that the scattered muons have to be hard, with $P_T> 25$ GeV.

\section{LHeC sensitivity to cLFV from heavy neutral leptons}

In this section, before we turn to the model-independent analysis, we investigate the sensitivity of the LHeC for cLFV induced by heavy neutral leptons (also referred to as ``heavy neutrinos'' or ``sterile neutrinos''). In particular, we will explore the LHeC sensitivity to the combinations $|\theta_e\theta^\ast_\mu|$ and $|\theta_e\theta^\ast_\tau|$ of active-sterile neutrino mixing angles within the "Symmetry Protected Seesaw Scenario'' (SPSS) benchmark scenario (cf.\ \cite{Antusch:2015mia,Antusch:2016ejd}), and compare it with the current bounds from non-collider experiments. The most relevant present constraints on the mixing parameters come from the two body decays, e.g. $\ell_\alpha\to \ell_e \gamma$ \cite{TheMEG:2016wtm,Aubert:2009ag}, and the three body decays $\ell_\alpha\to 3 \,\ell_e$ \cite{Bellgardt:1987du,Baranov:1990uh,Hayasaka:2010np} of taus and muons ($\alpha = \mu,\tau$). For final state muons we also consider the constraint from the $\mu - e$ conversion search at SINDRUM II \cite{Dohmen:1993mp}.

\subsection{Benchmark scenario: SPSS}\label{sec:SPSS}
For the analysis of the LHeC sensitivities and the comparison to the present experimental constraints, we consider the SPSS benchmark model. In this subsection, we will only give a brief summary to the SPSS and refer for details to \cite{Antusch:2015mia,Antusch:2016ejd}. Beyond the particle content of the SM, the scenario includes two sterile neutrinos with opposite charges under an approximate "lepton number''-like symmetry. The small observed neutrino masses arise from the small breaking of the "lepton number''-like symmetry. 
For the study of cLFV, we can treat the protective "lepton number''-like symmetry as being exact, such that lepton number is conserved. A discussion for which parameter regions the lepton number violating effects can be observable in the SPSS benchmark model with small symmetry breaking can be found in \cite{Antusch:2017ebe}. 

The Lagrangian density of the SPSS benchmark model, including the sterile neutrino pair $N_R^1$ and $N_R^2$, is given by:
\begin{equation}
\mathcal{L} = \mathcal{L}_\mathrm{SM} -  \overline{N_R^1}
M_N
N^{2\,c}_R - y_{\nu_{\alpha}}\overline{N_{R}^1} \widetilde \phi^\dagger \, L^\alpha
+\mathrm{H.c.}
+ \dots  \;,
\label{eqn:lagrange}
\end{equation}
where $L^\alpha$ ($\alpha=e,\mu,\tau$) and $\phi$ are the lepton and Higgs doublets, respectively, and the parameters $y_{\nu_{\alpha}}$ denote the complex-valued neutrino Yukawa couplings. $M_N$ is the heavy neutral lepton (Majorana) mass parameter. 
The dots indicate additional terms which can be neglected in this study. They may contain additional heavy neutral leptons that are decoupled from collider phenomenology and indirect searches as well as the terms which slightly break the "lepton number''-like symmetry.

After electroweak symmetry breaking the neutral leptons (i.e.\ the active and sterile neutrinos) have a symmetric mass matrix, which can be diagonalized by a unitary 5 $\times$ 5 matrix $U$, cf.\ \cite{Antusch:2015mia}. 
The mass eigenstates are $\tilde n_j = \left(\nu_1,\nu_2,\nu_3,N_4,N_5\right)^T_j = U_{j \alpha}^{\dagger} n_\alpha$. They include the three light neutrinos (which are actually massless in the symmetry limit) and two heavy neutrinos with (in the symmetry limit) degenerate mass eigenvalues $M_N$.
The off-diagonal block of the mixing matrix $U$ governs the interactions of the heavy neutrinos. It can be quantified by the active-sterile neutrino mixing angles $\theta_\alpha$ related to the neutrino Yukawa couplings $y_{\nu_{\alpha}}$ via 
\begin{equation}
\theta_\alpha = \frac{y_{\nu_\alpha}^{*}}{\sqrt{2}}\frac{v_\mathrm{EW}}{M_N}\,, \qquad |\theta|^2 := \sum_{\alpha} |\theta_\alpha|^2\,,
\label{def:thetaa}
\end{equation}
where $v_\mathrm{EW} = 246.22$ GeV denotes the vacuum expectation value of the Higgs field.
Due to the mixing of the active and sterile neutrinos, the heavy neutrino mass eigenstates participate in the weak interactions as 
\begin{eqnarray}
j_\mu^\pm & \supset &  \frac{g}{2} \, \theta_\alpha \, \bar \ell_\alpha \, \gamma_\mu P_L \left(-\mathrm{i} N_4 + N_5 \right) + \text{H.c.} \,, \label{eqn:weakcurrent1}\\
j_\mu^0 & = & \frac{g}{2\,c_W} \sum\limits_{i,j=1}^5 \vartheta_{ij} \overline{ \tilde n_i} \gamma_\mu P_L \tilde n_j\,, \\
\mathcal{L}_{\rm Yuk.} & \supset & \frac{M_N}{v_\mathrm{EW}} \sum\limits_{i=1}^3 \left(\vartheta_{i4}^* \overline{N_4^c}+ \vartheta_{i5}^*\overline{N^c_5}\right) H\, \nu_i +\text{ H.c.}  \,.
\label{eqn:weakcurrent2}
\end{eqnarray}
$g$ is the weak coupling constant, $c_W$ the cosine of the Weinberg angle and $P_L = {1 \over 2}(1-\gamma^5)$ is the left-chiral projection operator. $H$ denotes the real scalar Higgs boson and  $\vartheta_{ij} :=  \sum_{\alpha=e,\mu,\tau} U^\dagger_{i\alpha}U_{\alpha j}^{}$.

Finally, we note that in the symmetry limit of the SPSS benchmark model, only the moduli $|\theta_{e}|$, $|\theta_{\mu}|$ and $|\theta_{\tau}|$ of the active-sterile mixing angles and the (w.l.o.g.\ real and positive) mass parameter $M_N$ are physical. 
Furthermore, we remark that via the relation 
$
|V_{ \alpha N}|^2 = | \theta_\alpha |^2 \:,
$
one can readily translate our results (which we will give in terms of the active-sterile neutrino mixing angles $\theta_\alpha$)  to the neutrino mixing matrix elements $V_{ \alpha N}$ often used in the literature.

\subsection{Calculation of the form factors for the cLFV operators}
\begin{figure}[t!]
\centering
\includegraphics[scale=0.48]{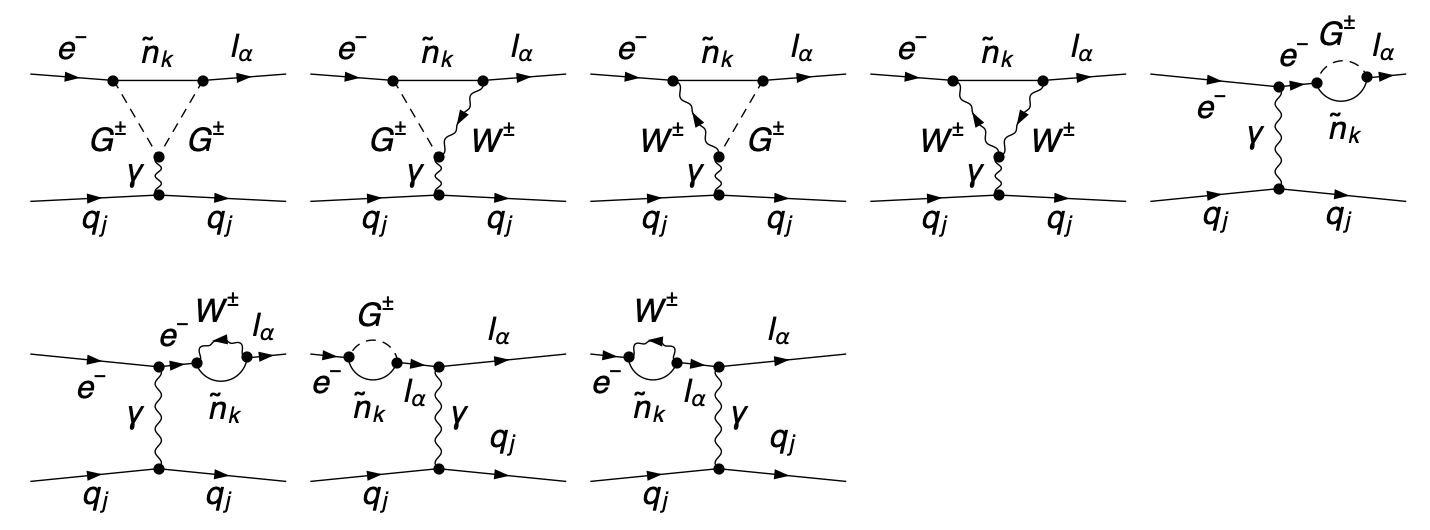}\\
\includegraphics[scale=0.48]{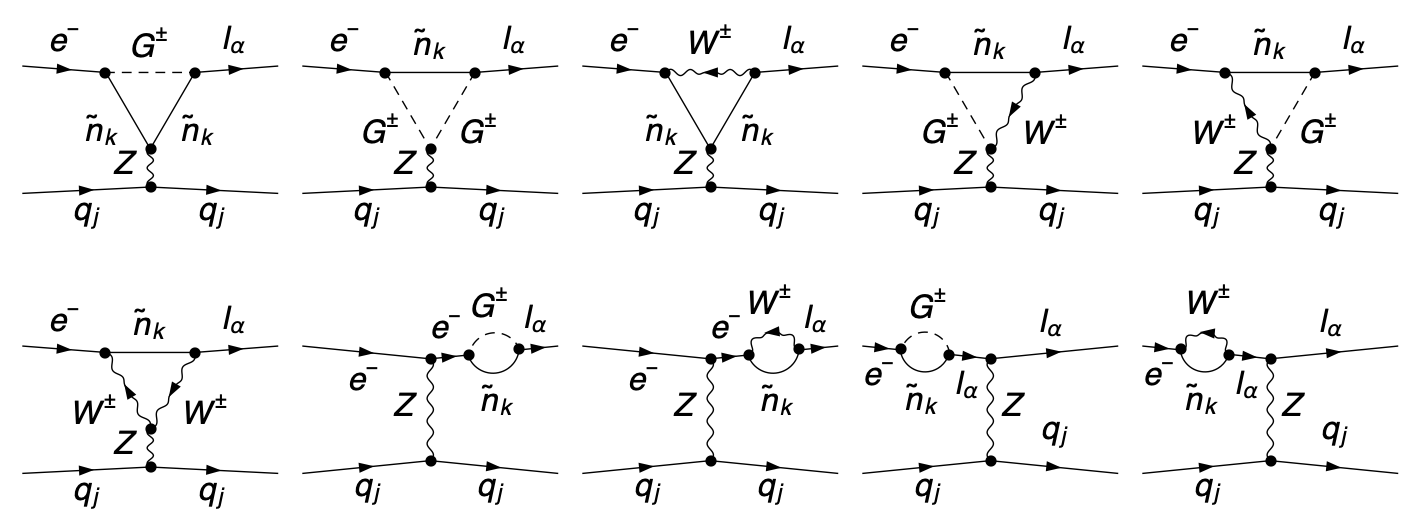}
\caption{Feynman diagrams generating the effective vertices for $e^- \to \ell_\alpha \gamma$ and $e^-\to \ell_\alpha Z$ in extensions of the SM by heavy neutral leptons. 
$\tilde{n}_k$ runs over all (light and heavy) neutral lepton mass eigenstates.
}
\label{fig:3}
\end{figure}

To calculate the form factors for the cLFV operators within the SPSS from the respective penguin diagrams (cf.\ Fig.~\ref{fig:3}), we use the package Peng4BSM@LO \cite{Bednyakov:2013tca}. Peng4BSM@LO is a Mathematica package that calculates the contributions of the form factors of certain effective operators originating from   one-loop penguin Feynman diagrams. In order to allow for generic finite form factors, the package calculates the form factors as the first order expansion of the small masses and momenta of the external fermions. We remark that all cLFV penguin processes have no tree level amplitude, and are thus finite at the one-loop level. The UV-divergence vanishes when we sum up over all diagrams and apply the unitarity condition of the leptonic mixing matrix $U$.

We find (using Peng4BSM@LO \cite{Bednyakov:2013tca}) that the form factors in the SM extension by heavy neutral leptons (within the SPSS benchmark scenario) are given by 
\begin{align*}
B^\gamma_L &= \sum_{k=1}^5\frac{e^2 |\theta_e\theta^\ast_{\alpha}|}{1152\pi^2 M^3_W\sin^3\theta_W(1-x^2_k)^4 }\left[ (x^2_k-1)(3e\ v_{EW} \ x^2_k(2x^4_k+5x^2_k-1)\right.\nonumber\\
&\left.-M_W\sin\theta_W(11x^6_k-27x^4_k+90x^4_k-20))
+12(M_W\sin\theta_Wx^4_k(x^4_k-4x^2_k+12)-3e\ v_{EW}\ x^6_k) \ln(x_k)\right],\\
\\
D^\gamma_L &= \frac{-i\ e^2\ M_e|\theta_e\theta^\ast_{\alpha}|}{384\pi^2 \  M^2_W\sin^2_W}\sum_{k=1}^5\left(\frac{7-34x^2_k+63x^4_k-34x^6_k-2x^8-(48x^6_k-12x^4_k)\ln(x_k)}{(1-x^2_k)^4}\right),\\
\\D^\gamma_R &= \frac{-i\ e^2\ M_{\alpha}|\theta_e\theta^\ast_{\alpha}|}{384\pi^2\  M^2_W\sin^2_W}\sum_{k=1}^5\left(\frac{7-34x^2_k+63x^4_k-34x^6_k-2x^8-(48x^6_k-12x^4_k)\ln(x_k)}{(1-x^2_k)^4}\right),\\
\end{align*}
\begin{align*}
A^Z_L &= \frac{ e^2|\theta_e\theta^\ast_{\alpha}|}{16\pi^2\ M_W\ \cos\theta_W\sin^3\theta_W }\\
&\times\sum_{k=1}^5\left(\frac{ 1 }{(1-x^2_k)^4}\left[  (x^2_k-1)(M_W(8x^2_k\sin^2\theta_W-9x^2_k-1)-4 e\sin\theta_W\ v_{EW} x^2_k)\right.\right. \nonumber\\
&\left.\left. +4(M_W(5-4\sin^2\theta_W)+2 e \sin\theta_W v_{EW})x^4_k \ln(x_k) \right]   \right),\\
\\
B^Z_L &= -\sum_{k=1}^5\frac{i \text{e}^2 |\theta_e\theta^\ast_{\alpha}|}{2304\pi^2 \cos\theta_W M^3_W \sin^3\theta_W (1-x^2_k)^4}\\
&\times \left[(x^2_k-1)(6 x^2_k \text{e } \sin\theta_W v_{EW} (2x^4_k+5x^2_k-1)+M_W(-12 -2 (\sin^2\theta_W-12)x^2_k\right.\nonumber\\ 
&\left.+(7\sin^2\theta_W-12)x^4_k-11x^6_k \sin^2\theta_W +\cos^2\theta_W(11x^6_k-47x^4_k+178x^2_k-40)))\right.\\
&\left. -12(6 x^2_k \text{e } \sin\theta_W v_{EW} x^6_k+M_W(4-14x^2_k+8(2+3 \cos^2\theta_W)x^4_k -2x^6_k(3+4 \cos^2\theta_W)\right.\\
&\left. +(\cos^2\theta_W-\sin^2\theta_W)x^8_k))\ln(x_k)\right],\\
\\
D^Z_L &= -\sum_{k=1}^5\frac{i \text{e}^2 M_e |\theta_e\theta^\ast_{\alpha}|}{768\pi^2 \cos\theta_W M^2_W \sin^3\theta_W (1-x^2_k)^4}\\
&\times \left[(x^2_k-1)(8+ x^2_k(\sin^2\theta_W-24) +x^4_k(16-5 \sin^2\theta_W) \right.\\
&\left. -2 x^6_k \sin^2\theta_W \cos^2\theta_W(14-53x^2_k+67x^4_k+2x^6)) -4x^2_k(2-2x^2_k(4+3 \cos^2\theta_W)\right.\\
&\left. +3 x^4_k(2+7\cos^2\theta_W-\sin^2\theta_W)\ln(x_k))\right],\\ 
\\
\end{align*}
\begin{align*}
D^Z_R &= -\sum_{k=1}^5\frac{i \text{e}^2 M_{\alpha} |\theta_e\theta^\ast_{\alpha}|}{768\pi^2 \cos\theta_W M^2_W \sin^3\theta_W (1-x^2_k)^4}\\
&\times \left[(x^2_k-1)(8+ x^2_k(\sin^2\theta_W-24) +x^4_k(16-5 \sin^2\theta_W)\right.\\
&\left. -2 x^6_k \sin^2\theta_W \cos^2\theta_W(14-53x^2_k+67x^4_k+2x^6)) \right.\\
&\left. -4x^2_k(2-2x^2_k(4+3 \cos^2\theta_W)+3 x^4_k(2+7\cos^2\theta_W-\sin^2\theta_W)\ln(x_k))\right],
\end{align*}
\begin{align*}
B^\gamma_R  = A^Z_R= B^Z_R = A^\gamma_{L,R} = 0\:.
\end{align*}
In the above equations, we have defined $x_k := \frac{M_{\tilde{n}_k}}{M_W}$. $e$ is the electric charge, $M_\alpha$ is the mass of the charged lepton $\ell_\alpha$ (with $\alpha = \mu,\tau$)  and $\theta_W$ denotes the weak mixing angle. 

We note that the lepton self energy diagrams with virtual photon exchange do not contribute to the amplitude since they cancel out with terms from $W$ boson and Goldstone boson diagrams. The  monopole term that is proportional to $q_\mu q_\nu$, cf.\ Eq.~(\ref{eq:2}), vanishes as it should because it would violate quark current conservation. 
For the case of $Z$ boson exchange the dipole form factors, $D^Z_{L,R}$,  flip the chirality of the outgoing fermions. They are suppressed since they are proportional to the lepton mass \cite{Bernabeu:1995sq,Budny:1976hq,Gutierrez-Rodriguez:2015rfa}.
 We have neglected the contributions from the effective operators with the SM Higgs boson, because they are suppressed by the small couplings of the Higgs to the beam quarks.

\subsection{Method for obtaining the cLFV sensitivity at the LHeC}

In the following, we assume that the heavy neutral leptons have sufficiently large masses that they cannot be directly produced at the LHeC. With this condition satisfied, we will  apply the effective operator treatment. 
The amplitudes for the $e-\mu$/$e-\tau$ conversion processes $pe^-\to \mu^- j$/$pe^-\to \tau^- j$
are given by 
\begin{align}
{\mathcal{M}_{LHeC}} = {\mathcal{M_{\gamma^\ast}}} + {\mathcal{M_{Z}}},
\end{align}
with $ {\mathcal{M_{\gamma^\ast}}}$ and $ {\mathcal{M_{Z}}}$ denoting the amplitudes for virtual photon and $Z$ boson exchange
\begin{align*}
&{\mathcal{M}_{\gamma^\ast}} = \bar{u}_{l_\alpha}\left[ B^\gamma_{L,R} P_{L,R} \,q^2\,\gamma^\nu - i\sigma^{\mu\nu}  q_\mu \, D^\gamma_{L,R} P_{L,R}  \right] u_{e} \left(\frac{-ie\ g_{\mu\nu}}{q^2} \right) \,\bar{u}_{q}(- ie Q_q\gamma^\mu) v_{q} \:, \\
&{\mathcal{M}_{Z}} = 
\bar{u}_{l_\alpha}\left[A^Z_{L,R}P_{L,R}\gamma^\nu +B^Z_{L,R}P_{L,R}q^2\gamma^\nu- i\sigma^{\mu\nu}  q_\mu D^Z_{L,R}P_{L,R}\right] u_{e} \left(\frac{-ig_{\mu\nu}}{q^2-M^2_Z} \right)\bar{u}_{q} (\gamma^\mu \,g_{L,R} P_{L,R}) v_{q}. 
\end{align*}
$Q_q$ is the quark charge and $g_{L,R}$ are the left and right couplings of the $Z$ boson with quarks (where again expressions like $g_{L,R} P_{L,R}$ stand for the sums, i.e.\ $g_{L} P_{L}+g_{R} P_{R}$). $B^\gamma_{L,R},D^\gamma_{L,R}, A^Z_{L,R}, B^Z_{L,R}$  and $D^Z_{L,R}$ are the effective form factors of the one-loop penguin diagrams in Fig.~\ref{fig:3}, with results given in the previous subsection.
 
In the following we will carry out the cLFV sensitivity analysis for the case of muons in the final state, $pe^-\to \mu^- j$, and taus in the final state, $pe^-\to \tau^- j$, separately. These two searches at the LHeC can test the combinations  $|\theta_e\theta^\ast_\mu|$ and $|\theta_e\theta^\ast_\tau|$ of the flavor-dependent active-sterile mixing angles, respectively, for a given heavy neutrino mass $M_N$.  
In the analysis with muons in the final state, we initially fix $|\theta_e\theta^\ast_\mu|= 10^{-3}$ with $\theta_e=\theta_\mu$ and $\theta_\tau=0$, and for the analysis  with taus in the final state we fix $|\theta_e\theta^\ast_\tau|= 10^{-3}$ with $\theta_e=\theta_\tau$ and $\theta_\mu=0$. We then use MadGraph \cite{Alwall:2014hca} to calculate the total cross section and generate the events, where the form factors and its Lorentz structure have been carefully implemented as described in \cite{Degrande:2011ua}. The parton shower and hadronisation are done by Pythia \cite{Sjostrand:2006za}. For fast detector simulation we use Delphes \cite{deFavereau:2013fsa}. For event reconstruction and analysis we use MadAnalysis \cite{Conte:2012fm,Conte:2014zja}.

\subsection{Event reconstruction and analysis}

For signal reconstruction (at the reconstructed level after detector simulation), we require at least one muon with $P_T \ge 25$ GeV and  jets with $P_T\ge 5$ GeV. For tau lepton reconstruction, we use an identification efficiency rate of $75\%$ for tau leptons with $P_T \ge 25$ GeV and misidentification rate about $1\%$ \cite{Bagliesi:2007qx,Bagliesi:2006ck}. For the process with final state taus,  we use Pythia for tau decays and then we use the Delphes analysis module to reconstruct the hadronic tau jet with identification efficiency rate of $75\%$  for the signal and identification efficiency rate of $60\%$ for the background. The most relevant backgrounds and their total cross sections are shown in Table \ref{tab:1} for final state taus (left) and final state muons (right).
\begin{table}[h!]
\resizebox{7.4cm}{!}{
\parbox{.68\linewidth}{
\begin{tabular}{|c|c|c|}
\hline
$\#$&Backgrounds $\tau$ final state & $\sigma_{(LHeC)} [Pb]$   \\ \hline\hline
bkg1&$p e^-\to Z/\gamma^\ast (\to \tau^-\tau^+)\ \nu_l\  j  $  & 0.0316   \\ \hline
bkg2&$p e^-\to  W^\pm(\to \tau^\pm\ {\nu}_\tau) \ e^-\  j $  &0.2657  \\ \hline
bkg3&$p e^-\to Z Z(\to \tau^-\tau^+) \ \nu_l\  j  $  & 1.1$\times 10^{-5}$   \\ \hline
bkg4&$p e^-\to Z(\to \tau^-\tau^+) W^\pm(\to \tau^\pm\ {\nu}_\tau) \ \nu_l\  j $&    2.64$\times 10^{-5}$   \\ \hline 
\end{tabular}
}}
\hfill
\resizebox{7.4cm}{!}{
\parbox{.68\linewidth}{
\begin{tabular}{|c|c|c|}
\hline
$\#$ &Backgrounds $\mu$ final state& $\sigma_{(LHeC)} [Pb]$   \\ \hline\hline
bkg1&$p e^-\to Z/\gamma^\ast (\to \mu^-\mu^+)\ \nu_l\  j  $  & 0.0316   \\ \hline
bkg2&$p e^-\to  W^\pm(\to \mu^\pm\ {\nu}_\mu) \ e^-\  j $  &0.2657  \\ \hline
bkg3&$p e^-\to Z/\gamma^\ast (\to \tau^-\tau^+\to\text{leptons})\ \nu_l\  j   $  &9.1$\times 10^{-4}$   \\ \hline
bkg4&$p e^-\to  W^\pm(\to \tau^\pm\ {\nu}_\tau\to\text{leptons}) \ e^-\  j $  & 0.0451   \\ \hline
bkg5&$p e^-\to Z Z(\to \mu^-\mu^+) \ \nu_l\  j  $  & 1.1$\times 10^{-5}$   \\ \hline
bkg6&$p e^-\to Z(\to \mu^-\mu^+) W^\pm(\to \mu^\pm\ {\nu}_\mu) \ \nu_l\  j $&    2.64$\times 10^{-5}$ \\\hline  
\end{tabular}
}}
\caption{Dominant background processes considered in our analysis and their total cross sections for final state taus (left) and final state muons (right). The cross sections  are obtained from MadGraph, while for the later tau decays we utilize Pythia.  The samples have been produced with the following parton level cuts: $P_T(j)\ge 5$ GeV, $P_T(l)\ge 2$ GeV and $|\eta(l/j)|\le 4.5$. }
\label{tab:1}
\end{table}

It is worth mentioning that other backgrounds like $pe^-\to h\nu j$ with the SM Higgs decaying to a lepton pair are suppressed by the small Yukawa couplings, while the process of single top production $pe^-\to\nu t $ is suppressed by the small involved CKM mixing matrix element.

In order to enhance the signal-to-background rate, we reconstruct four variables that can distinguish between the signal and all relevant backgrounds. 
\begin{figure}[h!]
\includegraphics[scale=0.35]{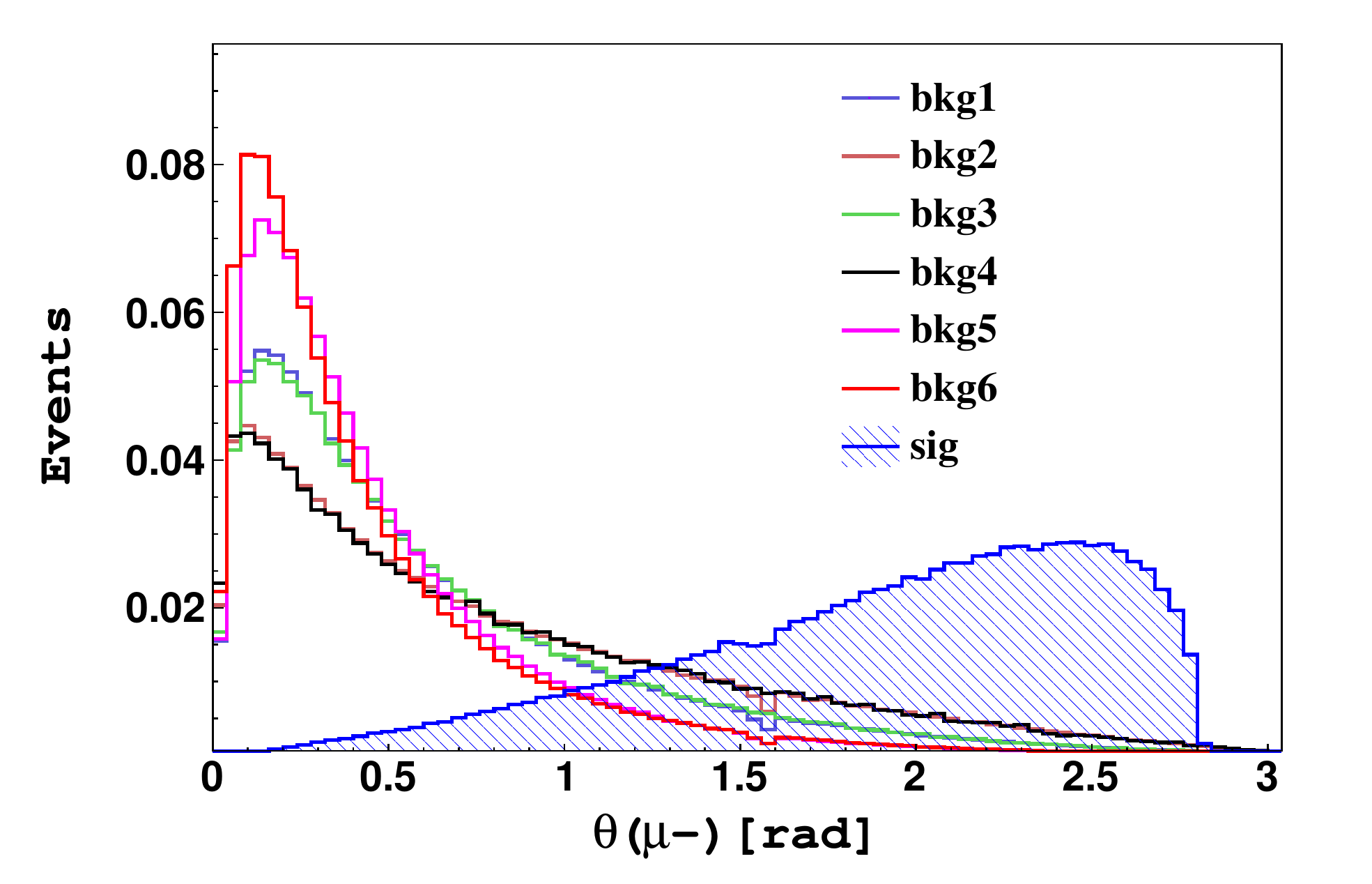}\includegraphics[scale=0.35]{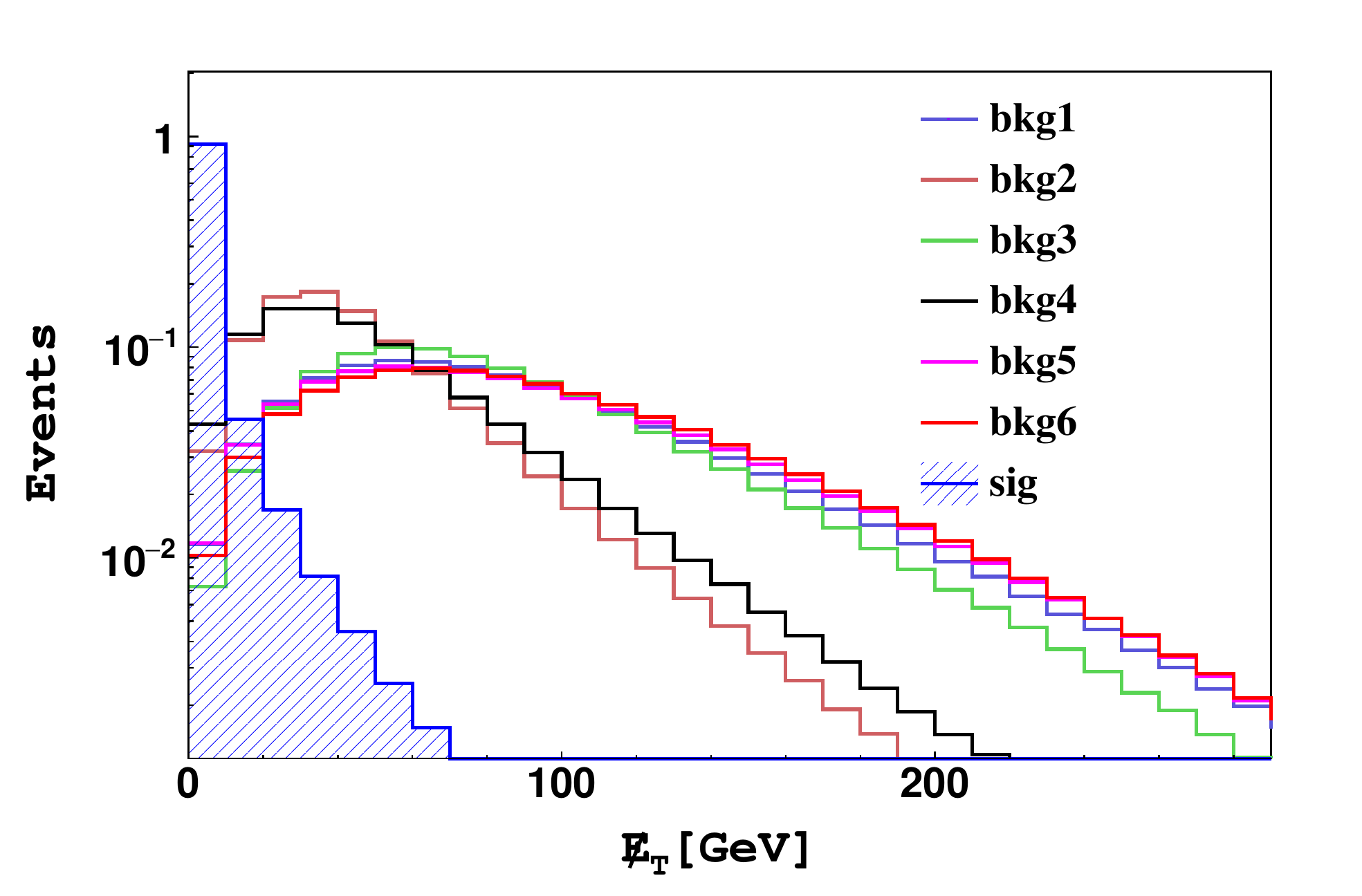}\\
\includegraphics[scale=0.35]{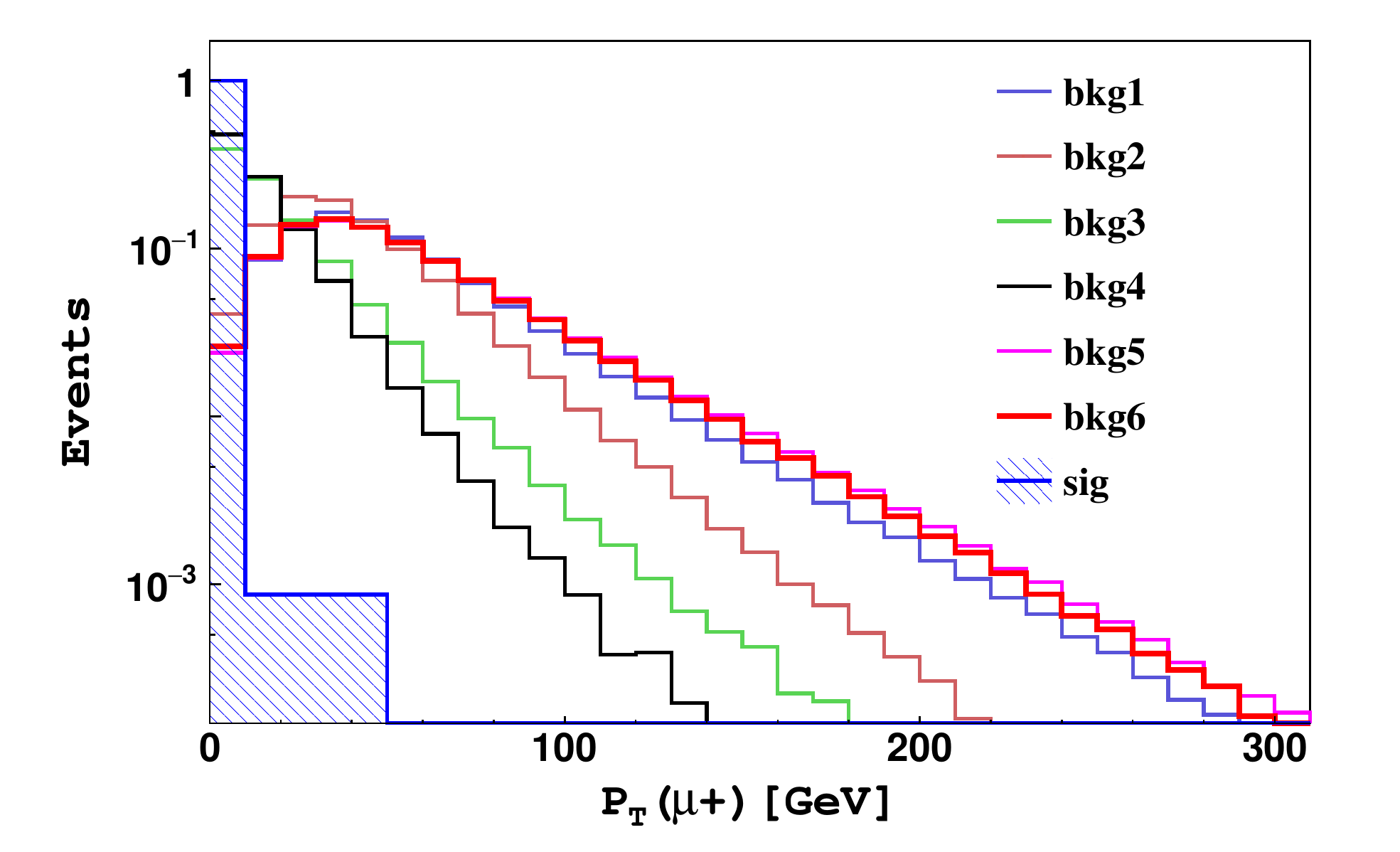}\includegraphics[scale=0.35]{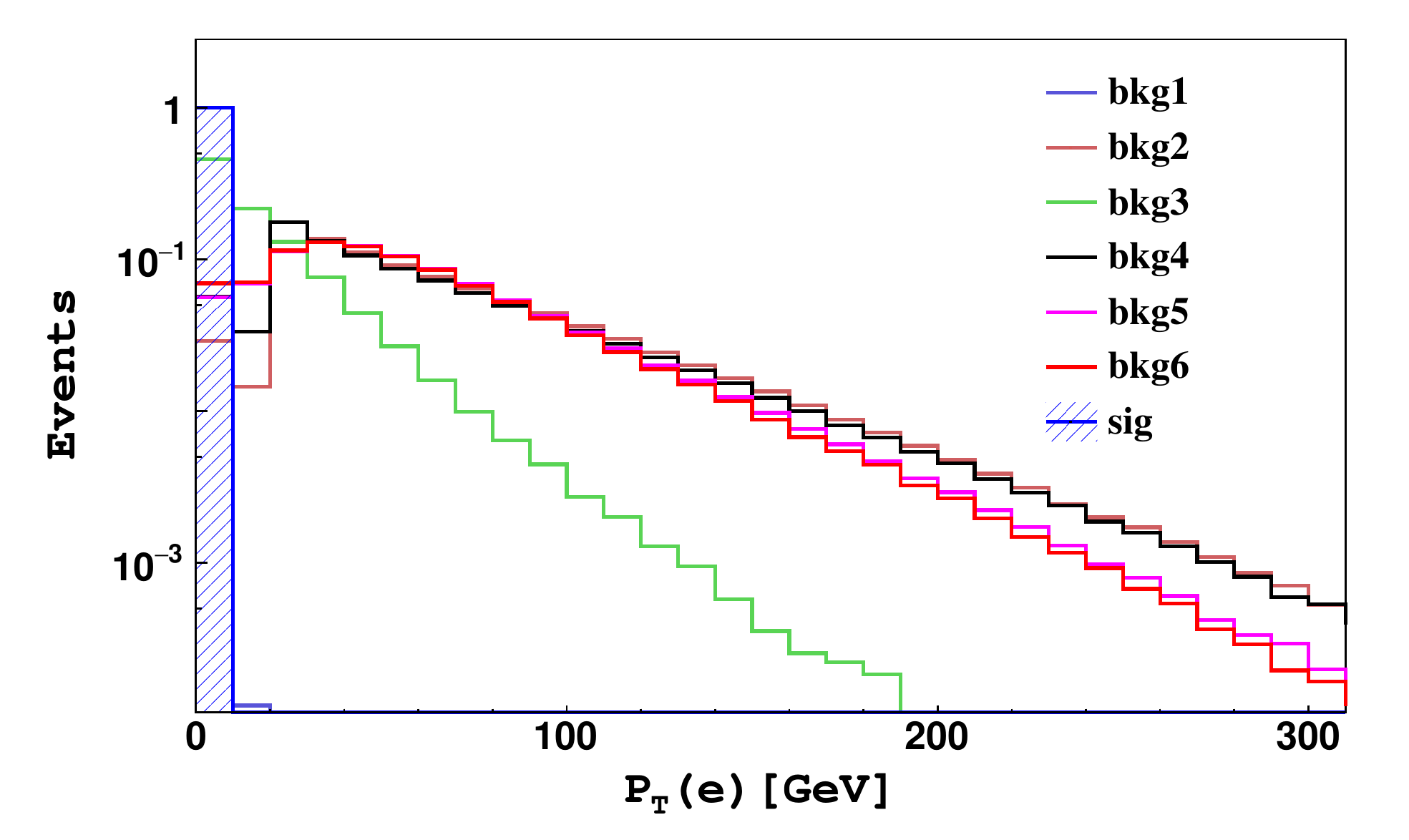}
\caption{Distributions of kinematic variables (before any cuts applied)  for the signal events with $M_N=1$ TeV, for the process $pe^-\to\mu^- j$ with muons in the final state, and with all relevant background events in Table \ref{tab:1} (right) superimposed and normalized to one. Upper left: angular distribution in radians for hard muons in the final state. Upper right: transverse missing energy. Down left: transverse momentum for anti-muons. Down right: transverse momentum for final state electrons.}
\label{fig:5}
\end{figure}
In Fig.~\ref{fig:5}, we show the kinematic distributions of the signal with final state muons versus all backgrounds superimposed.  The most important variable is the angular distribution of the final state hard leptons ($\mu/\tau$). They are mainly detected in the backward region of the detector while all the background processes produce hard leptons ($\mu/\tau$) in the forward region of the detector. For the case of signals with hard muons in the final state, the signal events have very low missing energy, while for taus in the final state there is a larger source of missing energy due to the hadronic tau reconstruction. Additionally, the transverse momenta of electrons or $\mu^+/\tau^+$ in the signal events are very small since the only source for them is the decay of radiated photons. In order to enhance the signal to background ratio, we optimize the cuts on these reconstructed kinematic variables as shown in Table \ref{tab:2} (left) for tau final states and (right) for muon final states for the benchmark point with $M_N=1$ TeV.

\begin{table}[h!]
\resizebox{7.4cm}{!}{
\parbox{.68\linewidth}{
\begin{tabular}{|c|c|c|}
\hline
Cut &Background events & Signal events   \\ \hline\hline
Normalized events (no cut)&297528&8147 \\ \hline
$P_T(\tau^+)\le 10$ GeV&  137986 &8117  \\ \hline
$\slashed{E}_T\le 100$ GeV&  132844 & 8110  \\ \hline
$P_T(e)\le 10$GeV& 14036&  8110  \\ \hline
$\theta(\tau^-) \ge 1.5$ rad&3561 &5302    \\ \hline
\end{tabular}
}}
\hfill
\resizebox{7.4cm}{!}{
\parbox{.68\linewidth}{
\begin{tabular}{|c|c|c|}
\hline
Cut &Background events & Signal events   \\ \hline\hline
Normalized events (no cut)&343600&11639 \\ \hline
$P_T(\mu^+)\le 10$ GeV&180114  &11596.75  \\ \hline
$\slashed{E}_T\le 50$ GeV&  126183 & 11517.4  \\ \hline
$P_T(e)\le 10$ GeV& 12705& 11517.3   \\ \hline
$\theta(\mu^-) \ge 1.5$ rad&4822.8 & 8925.9    \\ \hline
\end{tabular}}}
\caption{Cut efficiency, i.e.\ number of signal events and all backgrounds summed, for the processes $pe^-\to\tau^- j$ (left table) and $pe^-\to\mu^- j$ (right table) at the LHeC with integrated luminosity $3\ ab^{-1}$. For the signal events with final state taus we fix $\theta_e=\theta_\tau$, $\theta_\mu=0$ and  $|\theta_e\theta^\ast_\tau| = 10^{-3}$, which corresponds to a total cross section of $0.01173$ Pb (before the tau decays). For the signal events with muons in the final state we fix $\theta_e=\theta_\mu$, $\theta_\tau=0$ and  $|\theta_e\theta^\ast_\mu| = 10^{-3}$,  which corresponds to a total cross section of $0.01164$ Pb.  The heavy neutrino mass parameter $M_N$ has been set to $1$ TeV.  The numbers of signal and background events without cuts correspond to the above-given total cross sections and integrated luminosity.}
\label{tab:2}
\end{table}

\subsection{Results: sensitivities to the active-sterile mixing angles at the LHeC}

Given the number of signal events and the number of background events after the optimized cuts, the LHeC sensitivity at $90\%$ confidence level (CL) is obtained for  rejecting the signal plus background over the background-only hypothesis and by using the formula \cite{Antusch:2018bgr,Abe:2018bpo}
\begin{align}
\sigma_{sys} = \left[2\left((N_s+N_b) \ln\frac{(N_s+N_b)(N_b+\sigma_b^2)}{N^2_b+(N_s+N_b)\sigma^2_b} - \frac{N^2_b}{\sigma^2_b}\ln (1+\frac{\sigma^2_b N_s}{N_b(N_b+\sigma^2_b)} )\right)\right]^{1/2},
\label{eq:sigma}
\end{align}
with $N_s$ and $N_b$ being the number of signal and background events, and with $\sigma_b$ being the systematic uncertainty, taken to be $2\%$ \cite{AbelleiraFernandez:2012cc} for background events only.
For obtaining the current limits from non-collider experiments we use the following experimental constraints at $90\%$ CL:
\begin{align*}
 &Br(\mu\to e\gamma)\le 4.2\times 10^{-13}\   \cite{TheMEG:2016wtm} \:,  \\
 &Br(\tau\to e\gamma)\le 3.3\times 10^{-8}\   \cite{Aubert:2009ag} \:, \\
 &Br(\mu\to e^-e^+e^-)\le 1.\times 10^{-12}\   \cite{Bellgardt:1987du,Baranov:1990uh} \:, \\
 &Br(\tau\to e^-e^+e^-)\le 2.7\times 10^{-8}\   \cite{Hayasaka:2010np} \:, \\
 &Cr(\mu-e,\:^{197}_{79}\text{Au})\le 7\times 10^{-13}\    \text{\cite{Bertl:2006up}} \:.
\end{align*}
\begin{figure}[h!]
\centering
\includegraphics[scale=0.28]{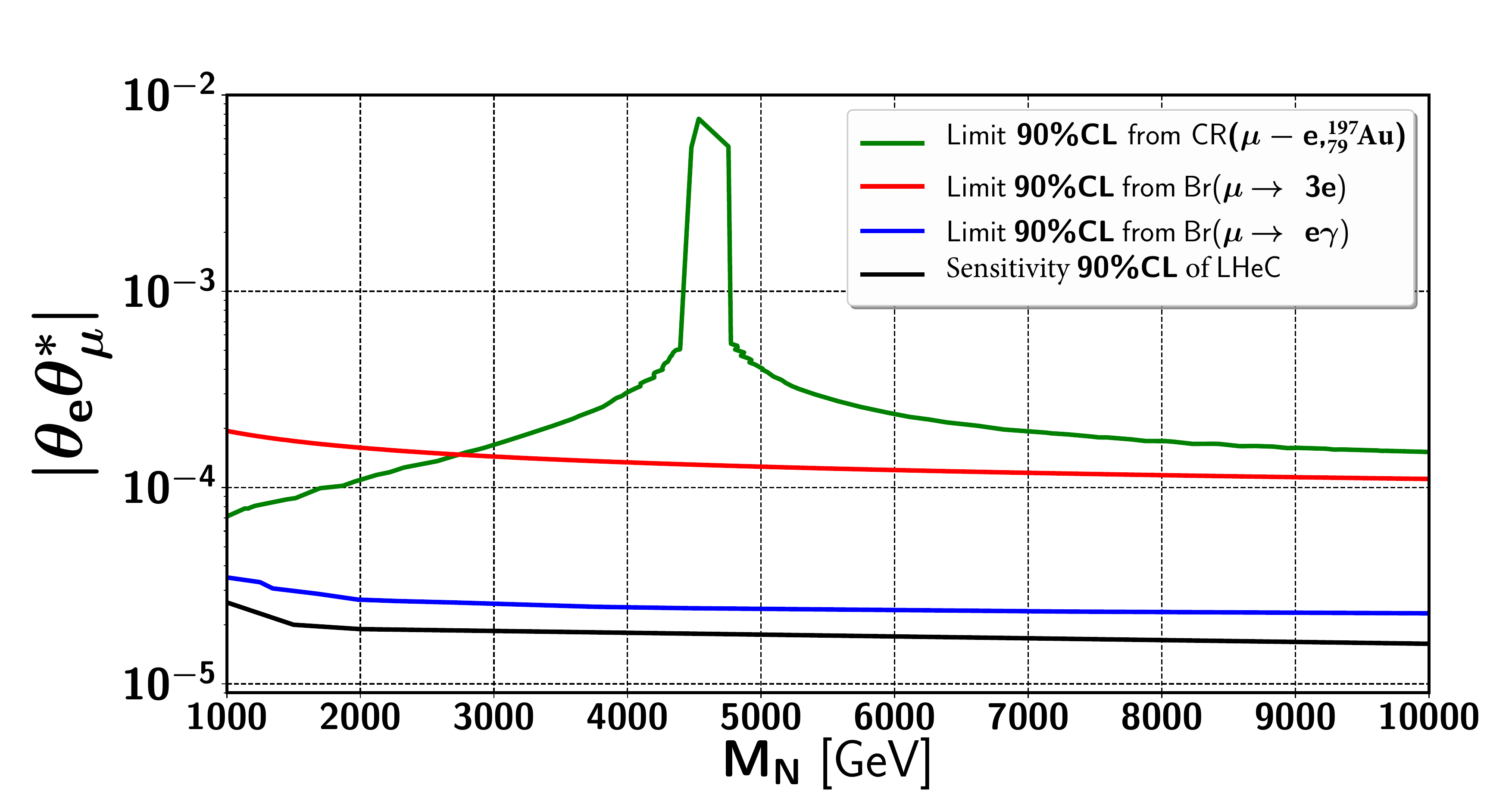}\\
\includegraphics[scale=0.28]{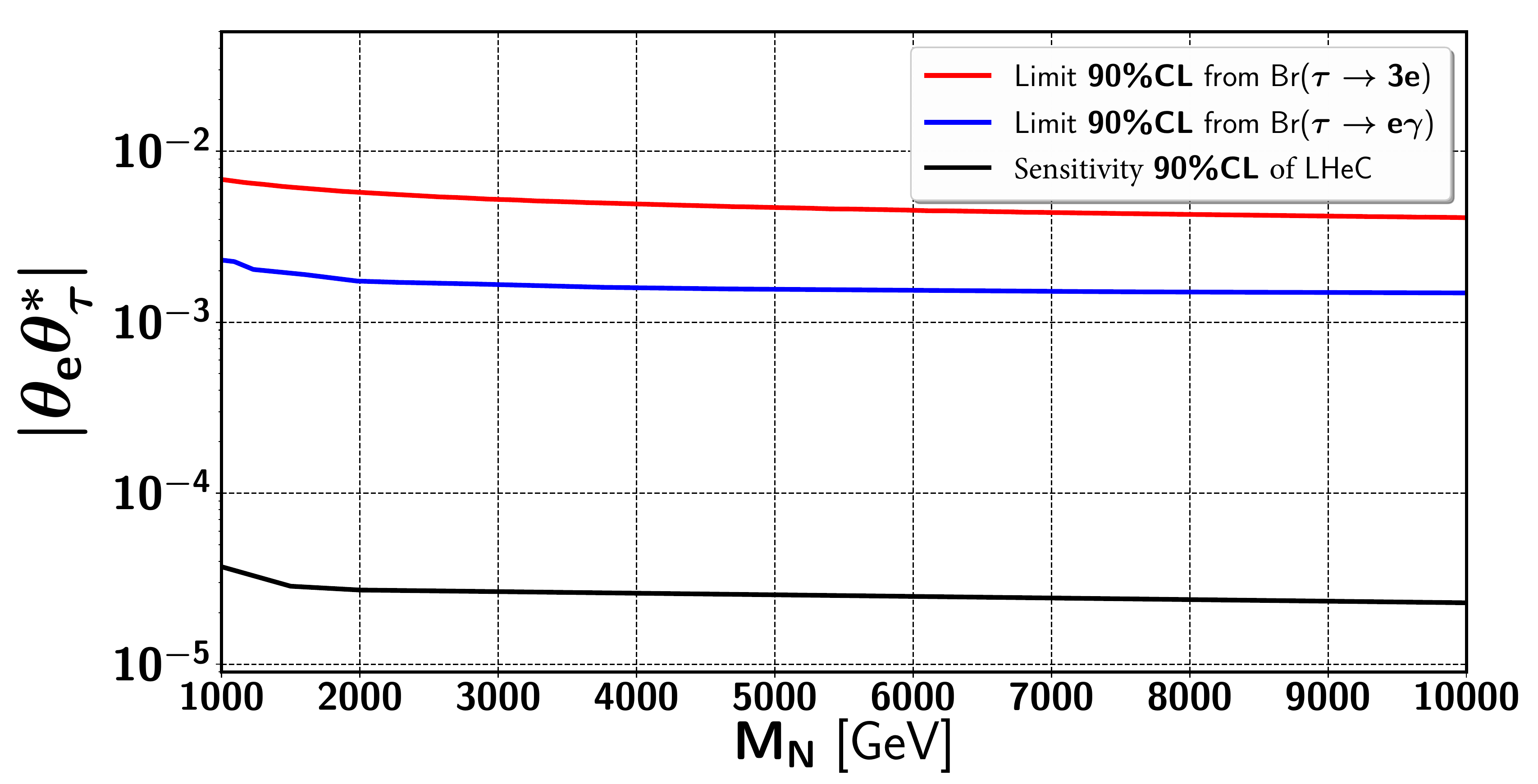}
\caption{ Estimated sensitivities to the active-sterile neutrino mixing angle combinations $|\theta_e\theta^\ast_\mu|$ (upper panel) and $|\theta_e\theta^\ast_\tau|$ (lower panel). The black lines show our results for the LHeC sensitivity from the processes $pe^-\to\mu^- j$ and $pe^-\to\tau^- j$, respectively, with $1.3$ TeV center-of-mass energy and integrated luminosity of $3\ \text{ab}^{-1}$. The green line in the upper panel corresponds the current limit from $\mu-e$ conversion, the red and blue lines in both panels show the current limits from $\ell_\alpha \to 3e$ and  $\ell_\alpha \rightarrow e \gamma$ (taken  from \cite{Alonso:2012ji}), respectively. }
\label{fig:limit}
\end{figure}

From the bounds on the branching (or conversion) ratios for $\ell_\alpha\to e\gamma$, $\ell_\alpha\to 3e$ and $\mu-e$ conversion in nuclei, we calculate the limits on the active-sterile neutrino mixing angles using the formulae given in \cite{Alonso:2012ji}. It is worth mentioning that the processes $\ell_\alpha\to 3e$ and $\mu-e $ conversion in nuclei have an energy scale of $q^2= M^2_\alpha$ (with $ \alpha =\mu,\tau$), which implies that the $Z$ boson contribution is suppressed by the squared mass difference in the propagator due to the small energy transfer \cite{Lee:1977tib}. On the other hand, at the LHeC the energy scale is $\sim 1.3$ TeV and thus the $Z$ boson can have a much larger contribution. The largest contribution indeed comes from the from the effective operator with form factor $A^Z_L$.

In Fig.~\ref{fig:limit} we present our results for the LHeC sensitivities to the active-sterile neutrino mixing angles and compare them with the current limits from non-collider experiments. The result with muons in the  final state (where the process is sensitive to $|\theta_e\theta^\ast_\mu|$) is shown in the upper plot, while the one with taus in the final state (sensitive to $|\theta_e\theta^\ast_\tau|$) is shown in the lower plot. The results show that with sensitivities down to $|\theta_e\theta^\ast_\mu| \le 2\times 10^{-5}$ and $|\theta_e\theta^\ast_\tau|\le 3\times 10^{-5}$ (for the example of $M_N = 1$ TeV), the LHeC can provide better sensitivity than the current  limit in both cases. 

Let us now also compare with the planned future experiments, using the sensitivity goals stated in section \ref{intro}. Regarding the mixing parameter combination $|\theta_e\theta_\mu^\ast|$, the future run of MEG II will improve the sensitivity to about $9\times 10^{-6}$, while the Mu3e experiment will be sensitive to this mixing parameter combination down to $4.8\times 10^{-6}$. The future PRISM even aims at a sensitivity of $2.5\times 10^{-6}$. These sensitivities would be better than the ones reachable at LHeC. 

Regarding the mixing parameter combination $|\theta_e\theta_\tau^\ast|$ responsible for the conversion of an electron into a tau, the sensitivity at the LHeC can be better than the current limits by more than an order of magnitude. Furthermore, the LHeC can even provide a better sensitivity than the future runs of the B-factories BABAR and BELLE II, which can probe the mixing parameter combination $|\theta_e\theta_\tau^\ast|$ down to $ 4\times 10^{-4}$ and $ 9\times 10^{-4}$, respectively.

Very sensitive test to the active-sterile neutrino mixing angles would also be possible at the FCC-ee \cite{Abada:2019lih}, via the accurate measurement of electroweak precision observables (EWPOs). On the other hand, the EWPOs are not sensitive to the same parameter combinations, but rather to $|\theta_e|^2+|\theta_\mu|^2$ and $|\theta_\tau|^2$. In the large mass limit for the heavy neutral leptons, sensitivities down to about $|\theta_e|^2+|\theta_\mu|^2 \sim 10^{-5}$ and $|\theta_\tau|^2 \sim 6\times 10^{-4}$ could be achieved \cite{Antusch:2016ejd,Antusch:2015mia,Antusch:2014woa}. We remark that e.g.\ for both $|\theta_e|$ and $|\theta_\tau|$ slightly below the (maximal) FCC-ee sensitivities from EWPOs, this would imply  $|\theta_e\theta_\tau^\ast| \lesssim 8\times 10^{-5}$, potentially still within reach of LHeC.

\section{Model-independent results}\label{sec:model-independent}
In this section, we calculate the model independent LHeC sensitivities for the form factors of the FCNC operators inducing cLFV given in section 2.1. The results can be used to estimate the LHeC discovery potential for generic heavy new physics that generates these effective operators. To calculate the LHeC sensitivities, we again analyse the processes $pe^-\to\mu^-j$ and $pe^-\to\tau^-j$, mediated by a cLFV effective coupling to photon, $Z$ boson, and SM Higgs. In the following, we use the form factors with superscripts to identify the considered boson, i.e.\ $\gamma$, $Z$ or $H$. To obtain the LHeC sensitivities,  we follow the same procedure used in the previous section and perform an analysis at the reconstructed level, switching on only one form factor at a time.
\begin{figure}[h!]
\centering
\includegraphics[scale=0.28]{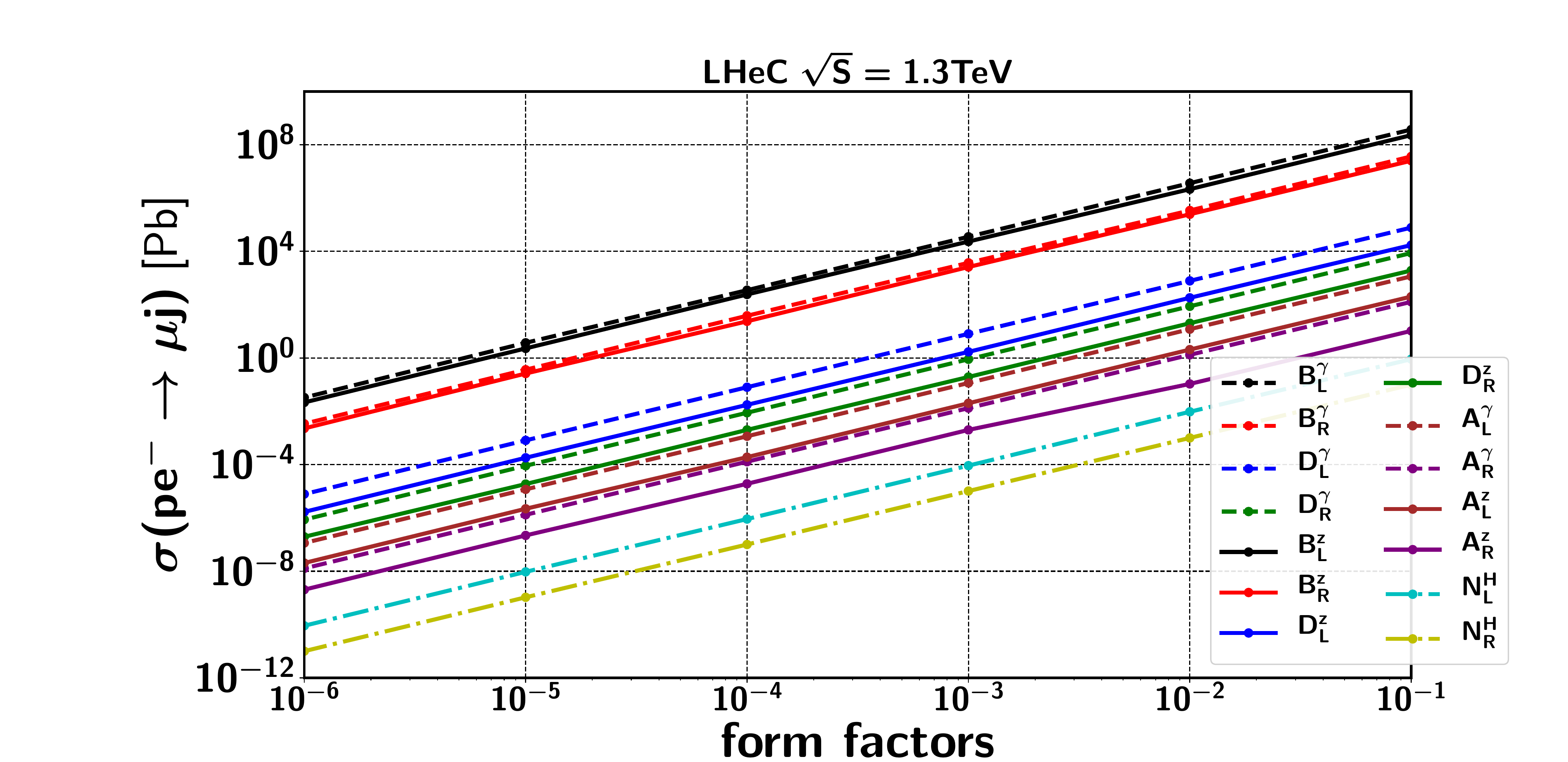}\\
\caption{Total cross section for the process $pe^-\to\mu j$ as function of the size of the individual form factors given in Eqs.~(\ref{eq:1}), (\ref{eq:2}) and (\ref{eq:3}), for the LHeC with $7$ TeV protons and $60$ GeV electrons with $80\%$ polarization. For the form factors $B^{Z/\gamma}_{L,R}$ and $D^{Z/\gamma}_{L,R}$, the $x$-axis shows their size in units of GeV$^{-2}$ and GeV$^{-1}$, respectively.
}
\label{fig:crosssection}
\end{figure}

In Fig.~\ref{fig:crosssection} we show the total cross section of $pe^-\to\mu^-j$ in picobarn as a function of the size of the individual form factors. One can see that the largest cross sections come from the  monopole form factors $B^{Z/\gamma}_{L,R}$,  which is due to the momentum transfer  squared attached to the effective vertex. The dipole form factors, $D^{Z/\gamma}_{L,R}$, also have comparatively large cross section due to the attached $q^\nu$ in the effective vertex. The form factors corresponding to the SM Higgs contribution have the lowest cross sections since the coupling of the SM Higgs with the proton beam is suppressed by the small Yukawa couplings. We remark that all considered kinematic distributions of the final state particles, except the angular distribution of the final state lepton, do not change by considering different form factors. 

\begin{figure}[h!] 
\centering
~~~\includegraphics[scale=0.6]{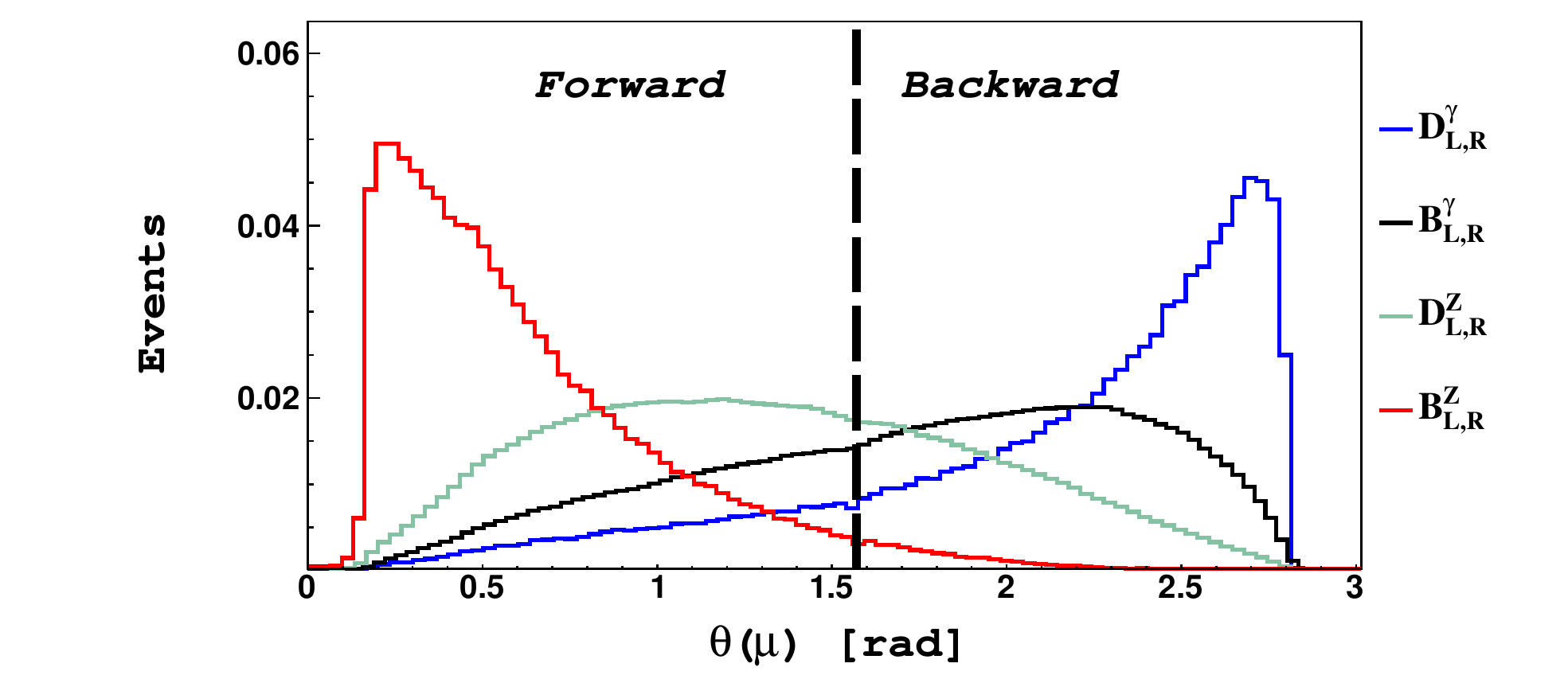}
\caption{Angular distribution of the muons for the process $pe^-\to\mu^- j$ at the reconstructed level, considering the monopole and dipole form factors for the effective operators that mediate the process via photon and Z boson exchange. The total event numbers are normalized to one. The forward direction is the proton beam  direction  and the backward direction is the electron beam direction.}
\label{fig:theta_FF}
\end{figure}

On the other hand, due to the dependence of the monopole and dipole form factors on the momentum of the mediator particle, the angular distributions of the final state leptons are shifted towards the forward direction. 
In Fig.~\ref{fig:theta_FF}, we show the angular distributions of the final state muons for the process $pe^-\to\mu^-j$, with total event number normalized to one. The shifting of the angular distributions towards the forward direction (in addition to the earlier discussed shifting for the processes with massive mediators compared to the photon-mediated processes) indeed weakens the signal vs.\ background separation, but still other characteristics such as $P_T(l^+), \slashed{E_T}$, and $P_T(e)$ can be used to improve the sensitivity.

\begin{table}[!htb]
\resizebox{\columnwidth}{!}{%
\begin{tabular}{|c|c|c|c|c|}
\hline
Form Factors&Cut &Background events & Signal events& LHeC sensitivity $90\%$  CL \\ \hline\hline
$N^H_L/N^H_R$&Normalized events (no cut)&343600&274/33& 4.49$\times 10^{-3}/3.55\times10^{-2}$\\ \hline
$N^H_L/N^H_R$&$P_T(\mu^+)\le 10$ GeV&180114  &274/33& 3.08$\times 10^{-3}/2.57\times 10^{-2}$ \\ \hline
$N^H_L/N^H_R$&$\slashed{E}_T\le 50$ GeV&  126183 & 269 /32& 2.65$\times 10^{-3}/2.22\times 10^{-2}$\\ \hline
$N^H_L/N^H_R$&$P_T(e)\le 10$GeV& 12705& 269/32  & 8.41$\times 10^{-4}/7.05\times10^{-3}$ \\ \hline
$N^H_L/N^H_R$&$\theta(\mu^-) \ge 0.5$ rad&9322 &232/28 &  8.36$\times 10^{-4}/6.90\times10^{-3}$  \\ \hline
\hline
Form Factors&Cut &Background events & Signal events&  LHeC sensitivity $90\%$  CL    \\ \hline\hline
$A^\gamma_L/A^\gamma_R$&Normalized events (no cut)&343600&$69000/7875$ &$1.75\times10^{-5}/1.50\times10^{-4}$\\ \hline
$A^\gamma_L/A^\gamma_R$&$P_T(\mu^+)\le 10$ GeV&  180114 &$68700/7840$  &$1.31\times10^{-5}/1.09\times10^{-4}$\\ \hline
$A^\gamma_L/A^\gamma_R$&$\slashed{E}_T\le 50$ GeV& 126183 &$68673/7837$ &$1.11\times10^{-5}/9.12\times10^{-5}$ \\ \hline
$A^\gamma_L/A^\gamma_R$&$P_T(e)\le 10$GeV& 12705&$68673/7837$  &$4.93\times10^{-6}/3.12\times10^{-5}$\\ \hline
$A^\gamma_L/A^\gamma_R$&$\theta(\mu^-) \ge 1.5$ rad&4823 &$67586/7713$& $3.94\times10^{-6}/2.16\times10^{-5}$  \\ \hline
\hline
Form Factors&Cut &Background events & Signal events&  LHeC sensitivity $90\%$  CL  \\ \hline\hline
$D^\gamma_L/D^\gamma_R\ \text{[GeV}^{-1}]$&Normalized events (no cut)&343600&2.41$\times10^7/2.66\times10^6$ &$1.58\times10^{-7}/1.42\times10^{-6}$\\ \hline
$D^\gamma_L/D^\gamma_R\ \text{[GeV}^{-1}]$&$P_T(\mu^+)\le 10$ GeV&  180114 &2.39$\times10^7/2.65\times10^6$  &$1.46\times10^{-7}/1.31\times10^{-6}$\\ \hline
$D^\gamma_L/D^\gamma_R\ \text{[GeV}^{-1}]$&$\slashed{E}_T\le 50$ GeV& 126183 &2.37$\times10^7/2.63\times10^6$ &$1.41\times10^{-7}/1.27\times10^{-6}$ \\ \hline
$D^\gamma_L/D^\gamma_R\ \text{[GeV}^{-1}]$&$P_T(e)\le 10$GeV& 12705& 2.37$\times10^7/2.63\times10^6$  &$1.12\times10^{-7}/1.01\times10^{-6}$\\ \hline
$D^\gamma_L/D^\gamma_R\ \text{[GeV}^{-1}]$&$\theta(\mu^-) \ge 0.3$ rad&10935 &2.37$\times10^7/2.63\times10^6$& $1.05\times10^{-7}/9.44\times10^{-7}$  \\ \hline
\hline
Form Factors&Cut &Background events & Signal events  & LHeC sensitivity $90\%$  CL  \\ \hline\hline
$B^\gamma_L/B^\gamma_R\ \text{[GeV}^{-2}]$&Normalized events (no cut)&343600&$9.52\times10^{10}/9.99\times10^{9}$&$1.35\times10^{-9}/1.28\times10^{-8}$ \\ \hline
$B^\gamma_L/B^\gamma_R\ \text{[GeV}^{-2}]$&$P_T(\mu^+)\le 10$ GeV&  180114 &$9.52\times10^{10}/9.99\times10^{9}$ &$1.31\times10^{-9}/1.25\times10^{-8}$ \\ \hline
$B^\gamma_L/B^\gamma_R\ \text{[GeV}^{-2}]$&$\slashed{E}_T\le 50$ GeV&  126183 & $9.26\times10^{10}/9.74\times10^{9}$&$1.31\times10^{-9}/1.25\times10^{-8}$  \\ \hline
$B^\gamma_L/B^\gamma_R\ \text{[GeV}^{-2}]$&$P_T(e)\le 10$GeV& 12705&  $9.26\times10^{10}/9.74\times10^{9}$&$1.21\times10^{-9}/1.15\times10^{-8}$ \\ \hline
$B^\gamma_L/B^\gamma_R\ \text{[GeV}^{-2}]$&$\theta(\mu^-) \ge 0.1$ rad&11898 &$9.25\times10^{10}/9.73\times10^{9}$&  $1.20\times10^{-9}/1.14\times10^{-8}$   \\ \hline
\hline
Form Factors&Cut &Background events & Signal events   & LHeC sensitivity $90\%$  CL \\ \hline\hline
$A^Z_L/A^Z_R$&Normalized events (no cut)&343600&17458/2182 &$6.77\times10^{-5}/5.37\times10^{-4}$\\ \hline
$A^Z_L/A^Z_R$&$P_T(\mu^+)\le 10$ GeV&  180114 &17394/2174  &$5.00\times10^{-5}/3.91\times10^{-4}$\\ \hline
$A^Z_L/A^Z_R$&$\slashed{E}_T\le 50$ GeV&  126183& 17276/2159 & $4.20\times10^{-5}/3.30\times10^{-4}$\\ \hline
$A^Z_L/A^Z_R$&$P_T(e)\le 10$GeV&  12705&  17276/2159&$1.54\times10^{-5}/1.07\times10^{-4}$ \\ \hline
$A^Z_L/A^Z_R$&$\theta(\mu^-) \ge 1.5$ rad&4823&13389/1674 & $1.36\times10^{-5}/8.74\times10^{-5}$  \\ \hline
\hline
Form Factors&Cut &Background events & Signal events  & LHeC sensitivity $90\%$  CL  \\ \hline\hline
$D^Z_L/D^Z_R\ \text{[GeV}^{-1}]$&Normalized events (no cut)&343600&$3.69\times10^6/4.22\times10^5$ &$5.66\times10^{-7}/4.95\times10^{-6}$\\ \hline
$D^Z_L/D^Z_R\ \text{[GeV}^{-1}]$&$P_T(\mu^+)\le 10$ GeV&  180114  &$3.68\times10^6/4.21\times10^5$ &$4.96\times10^{-7}/4.33\times10^{-6}$ \\ \hline
$D^Z_L/D^Z_R\ \text{[GeV}^{-1}]$&$\slashed{E}_T\le 50$ GeV&  126183 &$3.55\times10^6/4.06\times10^5$&$4.75\times10^{-7}/4.15\times10^{-6}$  \\ \hline
$D^Z_L/D^Z_R\ \text{[GeV}^{-1}]$&$P_T(e)\le 10$GeV&12705&$3.55\times10^6/4.06\times10^5$&$3.48\times10^{-7}/3.04\times10^{-6}$ \\ \hline
$D^Z_L/D^Z_R\ \text{[GeV}^{-1}]$&$\theta(\mu^-) \ge 0.1$ rad&11898 &$3.55\times10^6/4.06\times10^5$ &$3.45\times10^{-7}/3.01\times10^{-6}$   \\ \hline
\hline
Form Factors&Cut &Background events & Signal events  & LHeC sensitivity $90\%$  CL \\ \hline\hline
$B^Z_L/B^Z_R\ \text{[GeV}^{-2}]$&Normalized events (no cut)&343600&$6.99\times10^{10}/5.42\times10^9$ &$1.60\times10^{-9}/2.07\times10^{-8}$\\ \hline
$B^Z_L/B^Z_R\ \text{[GeV}^{-2}]$&$P_T(\mu^+)\le 10$ GeV&  180114 &$6.96\times10^{10}/5.39\times10^9$  &$1.56\times10^{-9}/2.01\times10^{-8}$\\ \hline
$B^Z_L/B^Z_R\ \text{[GeV}^{-2}]$&$\slashed{E}_T\le 50$ GeV&  126183 & $6.95\times10^{10}/5.39\times10^9$ & $1.53\times10^{-9}/1.97\times10^{-8}$\\ \hline
$B^Z_L/B^Z_R\ \text{[GeV}^{-2}]$&$P_T(e)\le 10$GeV&12705& $6.95\times10^{10}/5.39\times10^9$  &$1.41\times10^{-9}/1.82\times10^{-8}$\\ \hline

\end{tabular}}
\caption{LHeC sensitivities and cut efficiencies for the individual form factors (cf. section 2.1) of the FCNC operators inducing cLFV $e-\mu$ conversion, from the process $pe^-\to\mu^- j$  and with an integrated luminosity of $3 \text{ ab}^{-1}$.}
\label{tab:FF_mu}
\end{table}

\begin{table}[!htb]
\resizebox{\columnwidth}{!}{%
\begin{tabular}{|c|c|c|c|c|}
\hline
Form Factors&Cut &Background events & Signal events&  LHeC sensitivity $90\%$  CL \\ \hline\hline
$N^H_L/N^H_R$&Normalized events (no cut)&297528& $148/ 17$&$8.62\times10^{-3}/3.17\times10^{-2}$ \\ \hline
$N^H_L/N^H_R$&$P_T(\tau^+)\le 10$ GeV&  137986 & $148 / 17$&$5.12\times10^{-3}/2.12\times10^{-2}$ \\ \hline
$N^H_L/N^H_R$&$\slashed{E}_T\le 100$ GeV&  132844 & $147 /16$ &$5.09\times10^{-3}/2.01\times10^{-2}$ \\ \hline
$N^H_L/N^H_R$&$P_T(e)\le 10$ GeV& 14036&  $147 / 16$  & $1.61\times10^{-3}/1.48\times10^{-2}$\\ \hline
$N^H_L/N^H_R$&$\theta(\tau^-) \ge 0.5$ rad&8641 & $126 / 14 $ & $1.47\times10^{-3}/1.32\times10^{-2}$    \\ \hline
\hline
Form Factors&Cut &Background events & Signal events&  LHeC sensitivity $90\%$  CL \\ \hline\hline
$A^\gamma_L/A^\gamma_R$&Normalized events (no cut)&297528& $37260/4252$&$2.98\times10^{-5}/2.57\times10^{-4}$ \\ \hline
$A^\gamma_L/A^\gamma_R$&$P_T(\tau^+)\le 10$ GeV&  137986 & $37098/4234$&$2.09\times10^{-5}/1.76\times10^{-4}$ \\ \hline
$A^\gamma_L/A^\gamma_R$&$\slashed{E}_T\le 100$ GeV&  132844 & $37096/4234$ &$2.05\times10^{-5}/1.73\times10^{-4}$ \\ \hline
$A^\gamma_L/A^\gamma_R$&$P_T(e)\le 10$ GeV& 14036&  $37096/4234$  & $8.30\times10^{-6}/5.86\times10^{-5}$\\ \hline
$A^\gamma_L/A^\gamma_R$&$\theta(\tau^-) \ge 1.5$ rad&3561 & $36504 /4166 $ & $5.75\times10^{-6}/3.33\times10^{-5}$    \\ \hline
Form Factors&Cut &Background events & Signal events&   LHeC sensitivity $90\%$  CL \\ \hline\hline
$D^\gamma_L/D^\gamma_R\ \text{[GeV}^{-1}]$&Normalized events (no cut)&297528& $1.30\times10^7/1.44\times10^6$&$2.31\times10^{-7}/2.08\times10^{-6}$ \\ \hline
$D^\gamma_L/D^\gamma_R\ \text{[GeV}^{-1}]$&$P_T(\tau^+)\le 10$ GeV&  137986 & $1.29\times10^7/1.43\times10^6$&$2.07\times10^{-7}/1.86\times10^{-6}$ \\ \hline
$D^\gamma_L/D^\gamma_R\ \text{[GeV}^{-1}]$&$\slashed{E}_T\le 100$ GeV&  132844 & $1.29\times10^7/1.43\times10^6$ &$2.06\times10^{-7}/1.85\times10^{-6}$ \\ \hline
$D^\gamma_L/D^\gamma_R\ \text{[GeV}^{-1}]$&$P_T(e)\le 10$ GeV& 14036&  $1.29\times10^7/1.43\times10^6$  & $1.62\times10^{-7}/1.45\times10^{-6}$\\ \hline
$D^\gamma_L/D^\gamma_R\ \text{[GeV}^{-1}]$&$\theta(\tau^-) \ge 0.3$ rad&11993& $1.29\times10^7 /1.43\times10^6 $ & $1.61\times10^{-7}/1.45\times10^{-6}$    \\ \hline
Form Factors&Cut &Background events & Signal events&  LHeC sensitivity $90\%$  CL \\ \hline\hline
$B^\gamma_L/B^\gamma_R\ \text{[GeV}^{-2}]$&Normalized events (no cut)&297528& $5.14\times10^{10}/5.41\times10^9$&$1.88\times10^{-9}/1.79\times10^{-8}$ \\ \hline
$B^\gamma_L/B^\gamma_R\ \text{[GeV}^{-2}]$&$P_T(\tau^+)\le 10$ GeV&  137986 & $5.14\times10^{10}/5.41\times10^9$&$1.82\times10^{-9}/1.73\times10^{-8}$ \\ \hline
$B^\gamma_L/B^\gamma_R\ \text{[GeV}^{-2}]$&$\slashed{E}_T\le 100$ GeV&  132844 & $5.10\times10^{10}/5.43\times10^9$ &$1.81\times10^{-9}/1.72\times10^{-8}$ \\ \hline
$B^\gamma_L/B^\gamma_R\ \text{[GeV}^{-2}]$&$P_T(e)\le 10$ GeV& 14036&  $5.10\times10^{10}/5.43\times10^9$  & $1.67\times10^{-9}/1.59\times10^{-8}$\\ \hline
$B^\gamma_L/B^\gamma_R\ \text{[GeV}^{-2}]$&$\theta(\tau^-) \ge 0.1$ rad&12993& $5.10\times10^{10}/5.43\times10^9$ & $1.66\times10^{-9}/1.58\times10^{-8}$    \\ \hline
Form Factors&Cut &Background events & Signal events&  LHeC sensitivity $90\%$  CL \\ \hline\hline
$A^Z_L/A^Z_R$&Normalized events (no cut)&297528& $12221/1222$&$9.00\times10^{-5}/9.01\times10^{-4}$ \\ \hline
$A^Z_L/A^Z_R$&$P_T(\tau^+)\le 10$ GeV&  137986 & $12176/1218$&$6.19\times10^{-5}/6.18\times10^{-4}$ \\ \hline
$A^Z_L/A^Z_R$&$\slashed{E}_T\le 100$ GeV&  132844 & $12165/1217$ &$6.08\times10^{-5}/6.07\times10^{-4}$ \\ \hline
$A^Z_L/A^Z_R$&$P_T(e)\le 10$ GeV& 14036&  $12165/1217$  & $2.19\times10^{-5}/2.18\times10^{-4}$\\ \hline
$A^Z_L/A^Z_R$&$\theta(\tau^-) \ge 1.5$ rad&3561& $7953/795$ & $1.89\times10^{-5}/1.88\times10^{-4}$    \\ \hline

Form Factors&Cut &Background events & Signal events&   LHeC sensitivity $90\%$  CL \\ \hline\hline
$D^Z_L/D^Z_R\ \text{[GeV}^{-1}]$&Normalized events (no cut)&297528& $1.99\times10^6/2.28\times10^5$&$8.64\times10^{-7}/5.33\times10^{-6}$ \\ \hline
$D^Z_L/D^Z_R\ \text{[GeV}^{-1}]$&$P_T(\tau^+)\le 10$ GeV&  137986 & $1.98\times10^6/2.27\times10^5$&$7.24\times10^{-7}/3.95\times10^{-6}$ \\ \hline
$D^Z_L/D^Z_R\ \text{[GeV}^{-1}]$&$\slashed{E}_T\le 100$ GeV&  132844 & $1.97\times10^6/2.25\times10^5$ &$7.22\times10^{-7}/3.92\times10^{-6}$ \\ \hline
$D^Z_L/D^Z_R\ \text{[GeV}^{-1}]$&$P_T(e)\le 10$ GeV& 14036&  $1.97\times10^6/2.25\times10^5$  & $5.05\times10^{-7}/2.10\times10^{-6}$\\ \hline
$D^Z_L/D^Z_R\ \text{[GeV}^{-1}]$&$\theta(\tau^-) \ge 0.1$ rad&12993& $1.97\times10^6/2.25\times10^5$ & $5.00\times10^{-7}/2.07\times10^{-6}$    \\ \hline

Form Factors&Cut &Background events & Signal events&   LHeC sensitivity $90\%$  CL \\ \hline\hline
$B^Z_L/B^Z_R\ \text{[GeV}^{-2}]$&Normalized events (no cut)&297528& $3.78\times10^{10}/2.93\times10^9$&$2.22\times10^{-9}/9.13\times10^{-9}$ \\ \hline
$B^Z_L/B^Z_R\ \text{[GeV}^{-2}]$&$P_T(\tau^+)\le 10$ GeV&  137986 & $3.77\times10^{10}/2.91\times10^9$&$2.15\times10^{-9}/8.75\times10^{-9}$ \\ \hline
$B^Z_L/B^Z_R\ \text{[GeV}^{-2}]$&$\slashed{E}_T\le 100$ GeV&  132844 & $3.77\times10^{10}/2.91\times10^9$ &$2.18\times10^{-9}/8.88\times10^{-9}$ \\ \hline
$B^Z_L/B^Z_R\ \text{[GeV}^{-2}]$&$P_T(e)\le 10$ GeV& 14036&  $3.77\times10^{10}/2.91\times10^9$  & $2.00\times10^{-9}/7.94\times10^{-9}$\\ \hline
\end{tabular}}
\caption{LHeC sensitivities and cut efficiencies for the individual form factors (cf. section 2.1) of the FCNC operators inducing cLFV $e-\tau$ conversion,  from the process $pe^-\to\tau^- j$ and with an integrated luminosity of $3 \text{ ab}^{-1}$.}
\label{tab:FF_tau}
\end{table}

Our model-independent results are presented in Tables \ref{tab:FF_mu} and \ref{tab:FF_tau}, where we show the LHeC sensitivities to the individual form factors at 90$\%$ CL, based on the processes $pe^-\to\mu^-j$ and $pe^-\to\tau^-j$, respectively. For the analysis, we initially fix the values of the considered form factor to $10^{-3}$, with all other form factors set to zero, to calculate the total initial cross section which is used to normalize the generated events with integrated luminosity $3 \text{ ab}^{-1}$.  In order to increase the signal over background yield, the cuts have been optimized for each form factor individually. Given the number of signal and background events after each cut we have calculated the LHeC sensitivity at 90$\%$ CL for rejecting the signal plus background over the background-only hypothesis using the formula in Eq.~(\ref{eq:sigma}).

\section{Summary and conclusions}

In this work we have investigated the sensitivity of electron-proton ($ep$) colliders, in particular of the LHeC, for charged lepton flavor violation (cLFV). In an effective theory approach, we have considered a general effective Lagrangian for the conversion of an electron into a muon or a tau via the effective coupling of the charged leptons to a neutral gauge boson or a neutral scalar field. 

For the photon, the $Z$ boson and the Higgs particle of the SM, we have presented the sensitivities of the LHeC (with 3 $\text{ab}^{-1}$ integrated luminosity) for the coefficients of the effective operators (cf.\ section \ref{sec:model-independent} and Table \ref{tab:FF_mu} for the results with muons and Table \ref{tab:FF_tau} for the results with  taus in the final state), calculated from an analysis at the reconstructed level. 

As an example for a model where such flavor changing neutral current (FCNC) operators are generated at loop level, we have considered the extension of the Standard Model by sterile neutrinos in the context of the SPSS benchmark model. Our results for the sensitivities to the active-sterile neutrino mixing angle combinations $|\theta_e\theta^\ast_\mu|$ and $|\theta_e\theta^\ast_\tau|$ are shown in Fig.~\ref{fig:limit}.

Our results show that the LHeC (with 3 $\text{ab}^{-1}$ integrated luminosity) could already the LFV conversion of an electron into a muon beyond the current experimental bounds, and could reach more than an order of magnitude higher sensitivity than the present limits for the LFV conversion of an electron into a tau. 

We have argued that the very high sensitivities at the LHeC for some of the form factors are possible because the converted charged lepton is dominantly emitted in the backward direction, enabling an efficient separation of the signal from the background. The LHeC reach we obtained is in fact mainly statistics limited, and higher sensitivities could be achieved with higher integrated luminosity.

In summary, $ep$ colliders such as the proposed LHeC would be excellent facilities for probing cLFV. Especially for the case of cLFV electron-tau conversion, they could  reach the best sensitivities among all currently envisioned experiments, opening up a great discovery potential for new physics beyond the SM.

\section*{Acknowledgments}
This work has been supported by the Swiss National Science Foundation under the project number 200020/175502. A.R. acknowledges the hospitality of the Department of Physics, University of Basel, where the visit was supported through the SU-FPDC Grant Program. A.H.\ thanks Oliver Fischer for useful discussions.

\bibliographystyle{h-elsevier}
\bibliography{References}

\begin{thebibliography}{10}

\bibitem{TheMEG:2016wtm}
MEG, A. Baldini et~al.,
\newblock Eur. Phys. J. C 76 (2016) 434, 1605.05081.

\bibitem{Aubert:2009ag}
BaBar, B. Aubert et~al.,
\newblock Phys. Rev. Lett. 104 (2010) 021802, 0908.2381.

\bibitem{Hayasaka:2010np}
K. Hayasaka et~al.,
\newblock Phys. Lett. B 687 (2010) 139, 1001.3221.

\bibitem{Baldini:2018nnn}
MEG II, A. Baldini et~al.,
\newblock Eur. Phys. J. C 78 (2018) 380, 1801.04688.

\bibitem{Arndt:2020obb}
Mu3e, K. Arndt et~al.,
\newblock (2020), 2009.11690.

\bibitem{Miscetti:2020gkk}
Mu2e, S. Miscetti,
\newblock EPJ Web Conf. 234 (2020) 01010.

\bibitem{Shoukavy:2019ydh}
COMET, D. Shoukavy,
\newblock EPJ Web Conf. 212 (2019) 01006.

\bibitem{Alekou:2013eta}
A. Alekou et~al.,
\newblock {Community Summer Study 2013}: {Snowmass on the Mississippi}, 2013,
  1310.0804.

\bibitem{Kou:2018nap}
Belle-II, W. Altmannshofer et~al.,
\newblock PTEP 2019 (2019) 123C01, 1808.10567,
\newblock [Erratum: PTEP 2020, 029201 (2020)].

\bibitem{Bona:2007qt}
SuperB, M. Bona et~al.,
\newblock (2007), 0709.0451.

\bibitem{Lusiani:2010eg}
A. Lusiani,
\newblock PoS HQL2010 (2010) 054, 1012.3733.

\bibitem{Agostini:2020fmq}
LHeC, FCC-he Study Group, P. Agostini et~al.,
\newblock (2020), 2007.14491.

\bibitem{Bruening:2013bga}
O. Bruening and M. Klein,
\newblock Mod. Phys. Lett. A 28 (2013) 1330011, 1305.2090.

\bibitem{AbelleiraFernandez:2012cc}
LHeC Study Group, J. Abelleira~Fernandez et~al.,
\newblock J. Phys. G 39 (2012) 075001, 1206.2913.

\bibitem{Klein:2009qt}
M. Klein,
\newblock {17th International Workshop on Deep-Inelastic Scattering and Related
  Subjects}, p. 236, 2009, 0908.2877.

\bibitem{Antusch:2020fyz}
S. Antusch, A. Hammad and A. Rashed,
\newblock Phys. Lett. B 810 (2020) 135796, 2003.11091.

\bibitem{Jana:2019tdm}
S. Jana, N. Okada and D. Raut,
\newblock (2019), 1911.09037.

\bibitem{Flores-Sanchez:2019jcx}
O. Flores-S\'anchez et~al.,
\newblock PoS DIS2019 (2019) 094, 1908.09405.

\bibitem{Azuelos:2019bwg}
G. Azuelos et~al.,
\newblock Phys. Rev. D 101 (2020) 095015, 1912.03823.

\bibitem{Antusch:2019eiz}
S. Antusch, O. Fischer and A. Hammad,
\newblock JHEP 03 (2020) 110, 1908.02852.

\bibitem{DelleRose:2018ndz}
L. Delle~Rose, O. Fischer and A. Hammad,
\newblock Int. J. Mod. Phys. A 34 (2019) 1950127, 1809.04321.

\bibitem{Dev:2019hev}
P.S.B. Dev et~al.,
\newblock Phys. Rev. D 99 (2019) 115015, 1903.01431.

\bibitem{Klein:2008di}
M. Klein and R. Yoshida,
\newblock Prog. Part. Nucl. Phys. 61 (2008) 343, 0805.3334.

\bibitem{Alwall:2014hca}
J. Alwall et~al.,
\newblock JHEP 07 (2014) 079, 1405.0301.

\bibitem{Sjostrand:2006za}
T. Sjostrand, S. Mrenna and P.Z. Skands,
\newblock JHEP 05 (2006) 026, hep-ph/0603175.

\bibitem{deFavereau:2013fsa}
DELPHES 3, J. de~Favereau et~al.,
\newblock JHEP 02 (2014) 057, 1307.6346.

\bibitem{Conte:2012fm}
E. Conte, B. Fuks and G. Serret,
\newblock Comput. Phys. Commun. 184 (2013) 222, 1206.1599.

\bibitem{Antusch:2015mia}
S. Antusch and O. Fischer,
\newblock JHEP 05 (2015) 053, 1502.05915.

\bibitem{Antusch:2016ejd}
S. Antusch, E. Cazzato and O. Fischer,
\newblock Int. J. Mod. Phys. A 32 (2017) 1750078, 1612.02728.

\bibitem{Bellgardt:1987du}
SINDRUM, U. Bellgardt et~al.,
\newblock Nucl. Phys. B299 (1988) 1.

\bibitem{Baranov:1990uh}
V. Baranov et~al.,
\newblock Sov. J. Nucl. Phys. 53 (1991) 802.

\bibitem{Dohmen:1993mp}
SINDRUM II, C. Dohmen et~al.,
\newblock Phys. Lett. B317 (1993) 631.

\bibitem{Antusch:2017ebe}
S. Antusch, E. Cazzato and O. Fischer,
\newblock Mod. Phys. Lett. A 34 (2019) 1950061, 1709.03797.

\bibitem{Bednyakov:2013tca}
A.V. Bednyakov and c.H. Tany\i{}Ld\i{}Z\i{},
\newblock Int. J. Mod. Phys. C 26 (2014) 1550042, 1311.5546.

\bibitem{Bernabeu:1995sq}
J. Bernabeu, G. Gonzalez-Sprinberg and J. Vidal,
\newblock {Ringberg Workshop on Perspectives for Electroweak Interactions in e+
  e- Collisions}, pp. 0329--342, 1995, hep-ph/9505223.

\bibitem{Budny:1976hq}
R. Budny, B. Kayser and J. Primack,
\newblock Phys. Rev. D15 (1977) 1222.

\bibitem{Gutierrez-Rodriguez:2015rfa}
A. Guti\'errez-Rodr\'\i{}guez et~al.,
\newblock Nucl. Phys. B Proc. Suppl. 253-255 (2014) 202.

\bibitem{Degrande:2011ua}
C. Degrande et~al.,
\newblock Comput. Phys. Commun. 183 (2012) 1201, 1108.2040.

\bibitem{Conte:2014zja}
E. Conte et~al.,
\newblock Eur. Phys. J. C 74 (2014) 3103, 1405.3982.

\bibitem{Bagliesi:2007qx}
G. Bagliesi,
\newblock {17th Symposium on Hadron Collider Physics 2006 (HCP 2006)}, 2007,
  0707.0928.

\bibitem{Bagliesi:2006ck}
S. Gennai et~al.,
\newblock Eur. Phys. J. C 46S1 (2006) 1.

\bibitem{Antusch:2018bgr}
S. Antusch et~al.,
\newblock JHEP 10 (2018) 067, 1805.11400.

\bibitem{Abe:2018bpo}
LHC Dark Matter Working Group, T. Abe et~al.,
\newblock Phys. Dark Univ. 27 (2020) 100351, 1810.09420.

\bibitem{Bertl:2006up}
SINDRUM II, W.H. Bertl et~al.,
\newblock Eur. Phys. J. C 47 (2006) 337.

\bibitem{Alonso:2012ji}
R. Alonso et~al.,
\newblock JHEP 01 (2013) 118, 1209.2679.

\bibitem{Lee:1977tib}
B.W. Lee and R.E. Shrock,
\newblock Phys. Rev. D16 (1977) 1444.

\bibitem{Abada:2019lih}
FCC, A. Abada et~al.,
\newblock Eur. Phys. J. C 79 (2019) 474.

\bibitem{Antusch:2014woa}
S. Antusch and O. Fischer,
\newblock JHEP 10 (2014) 094, 1407.6607.

\end{thebibliography}
\end{document}